\documentclass[onecolumn, journal]{IEEEtran}
\usepackage[linesnumbered,ruled]{algorithm2e}
\usepackage{setspace}
\usepackage{cite}
\usepackage{amsmath,amssymb,amsfonts}

\usepackage{graphicx}
\usepackage{textcomp}
\usepackage{xcolor}

\usepackage[caption=false]{subfig}

\usepackage[colorlinks = true,
            linkcolor = blue,
            urlcolor  = blue,
            citecolor = blue,
            anchorcolor = blue]{hyperref}
\def\BibTeX{{\rm B\kern-.05em{\sc i\kern-.025em b}\kern-.08em
    T\kern-.1667em\lower.7ex\hbox{E}\kern-.125emX}}
\usepackage{float}
\usepackage{multirow}
\usepackage[noend]{algpseudocode}

\usepackage{adjustbox}

\usepackage{array}
\usepackage{url}

\begin{document}
%

\title{An Adaptive Learning based Generative Adversarial Network for One-To-One Voice Conversion}
%
%
%

\author{Sandipan Dhar,~\IEEEmembership{Student Member,~IEEE,}
    Nanda Dulal Jana,~\IEEEmembership{Member,~IEEE,} and Swagatam Das,~\IEEEmembership{}
        
\thanks{Sandipan Dhar is with the department of   computer science and engineering in the National Institute of Technology Durgapur, West Bengal, India (e-mail:{sd.19cs1101@phd.nitdgp.ac.in} ).}
\thanks{ Nanda Dulal Jana  is with the department of Computer Science and Engineering in the National Institute of Technology Durgapur, West Bengal, India (e-mail: nandadulal@cse.nitdgp.ac.in).}
\thanks{Swagatam Das is with the Electronics and Communication Sciences Unit, Indian Statistical Institute, Kolkata, India (e-mail:swagatam.das@isical.ac.in). }

}
\maketitle

\begin{abstract}
Voice Conversion (VC) emerged as a significant domain of research in the field of speech synthesis in recent years due to its emerging application in voice-assisting technology, automated movie dubbing, and speech-to-singing conversion to name a few. VC basically deals with the conversion of vocal style of one speaker to another speaker while keeping the linguistic contents unchanged. VC task is performed through a three-stage pipeline consisting of speech analysis, speech feature mapping, and speech reconstruction. Nowadays the Generative Adversarial Network (GAN) models are widely in use for speech feature mapping from source to target speaker. In this paper, we propose an adaptive learning-based GAN model called ALGAN-VC for an efficient one-to-one VC of speakers. Our ALGAN-VC framework consists of some approaches to improve the speech quality and voice similarity between source and target speakers. The model incorporates a Dense Residual Network (DRN) like architecture to the generator network for efficient speech feature learning, for source to target speech feature conversion. We also integrate an adaptive learning mechanism to compute the loss function for the proposed model. Moreover, we use a boosted learning rate approach to enhance the learning capability of the proposed model. The model is trained by using both forward and inverse mapping simultaneously for a one-to-one VC. The proposed model is tested on Voice Conversion Challenge (VCC) 2016, 2018, and 2020 datasets as well as on our self-prepared speech dataset, which has been recorded in Indian regional languages and in English. A subjective and objective evaluation of the generated speech samples indicated that the proposed model elegantly performed the voice conversion task by achieving high speaker similarity and adequate speech quality.
\end{abstract}


%
\IEEEpeerreviewmaketitle

\section{Introduction}
\begin{figure}[ht]
    \centering
    \captionsetup{justification=centering,margin=2cm}
    \includegraphics[height=0.17\linewidth, width=0.5\linewidth]{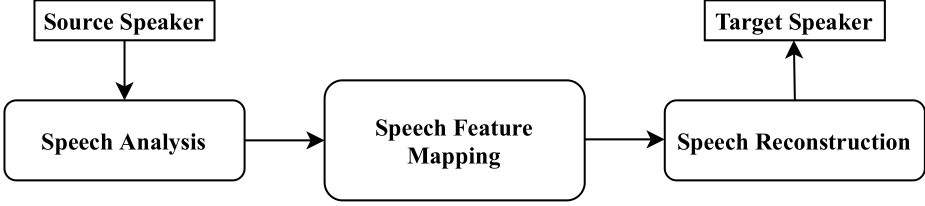}
    \caption{Workflow of traditional VC system}
    \label{Workflow-of-VC}
\end{figure}

%
%
%
%
\IEEEPARstart{I}{n} recent years Voice Conversion (VC) or Vocal Style Transfer (VST) is an emerging area of research in the field of speech synthesis. It is widely in use for many real-time applications such as audio assistive devices for speech disorder patients \cite{speech-disorder}, voice over in movie dubbing \cite{tv-dubbing}, speech-to-singing conversion \cite{singing-voice-conversion} etc. VC deals with the artificial generation of human speech by transferring the vocal tone  of source speaker to target speaker’s speech, without altering the content of the speech. Transformation of Vocal style from source speaker to target speaker is achieved by modifying speech features such as, fundamental frequency (${F}_{0}$), spectral envelop, formant structure etc for generating natural sounding speech. A traditional VC process involves mainly three components \cite{review-paper-1}: speech analysis, speech feature mapping and speech reconstruction. The VC components are illustrated in Fig. \ref{Workflow-of-VC}. The speech analysis component is responsible for decomposing the input speech sample into the features which are represented in the form of supra-segmental and segmental information. The speech feature mapping component treated as a mapping function from source speaker to target speaker's vocal features. This component plays an important role in any VC system for efficient mapping of speech characteristics from source speaker to target speaker. Finally, the speech reconstruction component is responsible for reconstruction of audible speech from the converted speech features. 

\par
VC systems are mainly divided into two categories based on conversion approach: Text-to-Speech (TTS) based VC system and Speech-to-Speech (STS) based VC system. Moreover, based on the speech datasets' linguistic contents and the speech feature mapping approach, a VC system can be classified into parallel and non-parallel VC-system. In parallel VC systems, training data consists of samples that contain similar linguistic contents (e.g., same spoken words or speech) of different speakers. Here, the speech feature mapping component can easily map a source speaker vocal features to a target speaker's vocal features because of the frame-wise alignment of the linguistic contents. Hence, parallel VC systems generate a high vocal style similarity score from listener’s perspective. On the other hand, in non-parallel VC systems,
training data consists of samples that contain misalignment of linguistic contents of source and target speaker's. Due to this misalignment of linguistic contents, it is hard to map a source speaker vocal features to a target speaker's vocal features as compared to parallel VC systems. Thus, parallel VC systems perform superior to non-parallel VC systems. However, in real-time applications, it is unexpected to have similar linguistic information of multiple speakers. Therefore, the non-parallel VC task is more realistic despite its difficulties.

\par
Most of the earlier VC models are developed on the basis of parametric and non-parametric statistical approaches \cite{speech-synthesis-review}. These approaches are strictly dependent on mono-lingual parallel data, which limits their use in real-time applications \cite{disadvantages-vc}. Moreover, some of the parallel data free statistical models often suffers from over-smoothing problems which degrade the quality of generated speech samples. With the recent advancements of Deep Learning (DL) techniques, speech-to-speech based VCs are gaining new momentum to overcome the limitations posed in statistical based VC models \cite{review-paper-1}.

\par
The progressive development of DL techniques opened up new research directions for DL-based generative models in VC applications \cite{speech-synthesis-review}. DL-based generative models or deep generative models' primary advantage is that these models are capable of learning significant features from large datasets. Generative Adversarial Network (GAN) \cite{gan-ian-goodfellow} models have emerged as an effective alternative for VC task among the deep generative models \cite{review-paper-1}. It has the potential ability to generate high-quality realistic data over speech data during vocal style transfer. Recently, many GAN-based VC models and their variants have been proposed in various articles \cite{deep-learning-1,deep-learning-2,deep-learning-3,deep-learning-5,deep-learning-4,deep-learning-6,deep-learning-7,deep-learning-8,deep-learning-9,deep-learning-10,deep-learning-12,deep-learning-13,deep-learning-14,deep-learning-15,cyclegan-vc-1}. The literature shows that a large section of GAN-based VC models are designed based on CycleGAN \cite{cyclegan-vc-1,cyclegan-vc-2,cyclegan-vc-3,cyclegan-vc-4,cyclegan-vc-5,cyclegan-vc-7,cyclegan-vc-8,cyclegan-vc-9,wave-cyclegan-2,childrens-gan,cyclegan-vc-100,cyclegan-vc-101,cyclegan-vc-102,cyclegan-vc-103} framework which is initially developed for image-to-image style translation. The published models have mainly experimented on mono-lingual parallel and non-parallel VC as compared to cross-lingual speech data. Moreover, their generated speech samples are not adequate in terms of speech quality and speaker similarity. Therefore, it is necessary to design an efficient and effective model for achieving high speech quality and speaker similarity concerning cross-lingual speech data.

\par
In this paper, we proposed an Adaptive Learning-based GAN model for one-to-one VC, called ALGAN-VC under speech-to-speech synthesis paradigm. A dense residual network (DRN) like architecture is incorporated in the generator network of the proposed model to enhance the feature learning ability of the model. An adaptive learning mechanism is proposed with an objective to derive a loss function which is formulated as a linear combination of Least Absolute Deviation ($L_1$) and Least Square Error ($L_2$) loss in our model. Both the $L_1$ and $L_2$ losses are computed by taking a minimum of the difference between the normalized output values of Mel-Cepsral Coefficient (MCEP) features and the target variables iteratively. The normalized MCEPs are determined with multiple activation functions such as ReLU, ELU, SELU, Leaky ReLU, and Sigmoid. Finally, a boosted learning rate strategy is used to enhance
the proposed ALGAN-VC model’s learning capability. The ALGAN-VC model is trained using both forward and inverse mapping simultaneously for one to one VC.

\par
The proposed model is experimented on Voice Conversion Challenge (VCC) 2016 \cite{vcc-2016-audio-data}, 2018 \cite{vcc-2018-audio-data}, 2020 \cite{vcc-2020-audio-data} datasets and our self prepared speech dataset which are recorded in English language as well as in Indian regional languages (Bengali, Hindi, Assamese, Odia, Malayalam). One-to-one speech-to-speech strategy is considered in our experiments for VC. Subjective and objective performance evaluations stated the proposed model's effectiveness in terms of speaker similarity and adequate speech quality. The Experimental results exhibited satisfactory improvement by transferring the vocal texture of a source speaker to a particular target speaker without changing the linguistic contents compared to state-of-the-art CycleGAN-based VC.
\par
The main contributions to the proposed ALGAN-VC model are summarized as follows:
\begin{enumerate}
\item{A dense residual network (DRN) like architecture is incorporated in the generator network of the proposed model to enhance the feature learning ability of the proposed model.}
\item{An adaptive learning approach is proposed for efficient feature learning concerning loss computation, and activation function is selected adaptively.}
\item{A boosted learning rate approach is used to enhance the proposed model’s learning capability.}
\item {A speech dataset is prepared by ourself which are recorded in the English language as well as in the Indian regional languages for the evaluation of the proposed model.}
\end{enumerate}

The rest of the paper is organized as follows. Section \ref{related-work} includes the overview of GAN model and related works on GAN-based VC models. The proposed ALGAN-VC model briefly presented in Section \ref{proposed-approach}. Section \ref{exp-setup} provides the details of the experimental setups and used speech datasets. Results and discussion with evaluation metrics are presented in Section \ref{result}. Finally, Section \ref{conclusion} includes conclusions and some promising future research directions on GAN based VC. 

\section{Background and Related Work}\label{related-work}
\subsection{Generative Adversarial Network (GAN)}
\begin{figure}[ht]
    \centering
    \includegraphics[height=0.19\linewidth, width=0.5\linewidth]{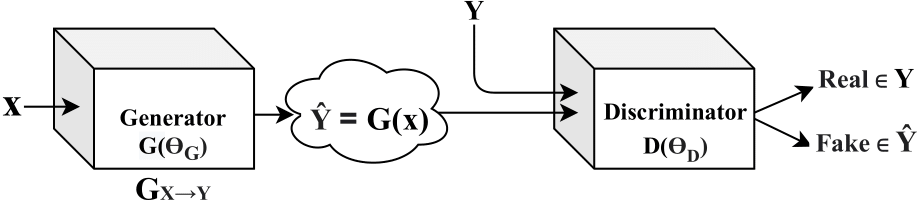}
    \caption{Basic framework of GAN}
    \label{Basic-framework-of-GAN}
\end{figure}

In recent years, deep generative models are widely used for synthetic data generation in various applications \cite{review-paper-1}. The GAN model is the most impressive and promising one among the deep generative models \cite{Review-paper-on-GAN}. A typical GAN model is consists of a generator and a discriminator which are denoted by $G$ and $D$. The generator generates synthetic data from a prior input information where the discriminator discriminates the generated synthetic data from the original training data. 
The basic framework of a GAN model is depicted in Fig. \ref{Basic-framework-of-GAN}.
The set of model parameters of a generator and a discriminator are represented by $\Theta_{G}$ and $\Theta_{D}$, respectively. Given an input data $\mathbf{x}$, the generator generates synthetic or fake data $\hat{y}$ with a probability distribution denoted as $p_{\hat{y}}$. On the other hand, the discriminator is trained with the original or real data $\mathbf{y}$ whose probability distribution is ${p_{\mathbf{y}}}$. Hence, the generator tries to match $\mathbf{x}$ with $\mathbf{y}$ using a mapping function which is represented by ${G_{\mathbf{x} \rightarrow \mathbf{y}}}$. For the real data $\mathbf{y}$ discriminator gives a probability value ${D_\mathbf{y}}$, which indicates the probability of belongingness of $\mathbf{y}$ to real data distribution. Similarly, for the generated fake data $\hat{y}$ the discriminator gives a probability value $D_{\hat{y}}$ which indicates the probability of belongingness of fake data to the real data distribution. The objective of the generator in GAN is to deceive the discriminator by generating real like synthetic data. On the other hand, discriminator's objective is to enhance its accuracy of classifying fake sample $\hat{y}$ as fake and real sample $\hat{y}$ as real. Thus, both the discriminator and the generator learns in an adversarial manner \cite{gan-ian-goodfellow} by maximizing and minimizing the objective function or loss function defined as
\begin{equation}
\begin{split}
\mathcal{L}_{CEL}(\mathbf{y},{\mathbf{\hat{y}}} )=\underset{G}{\mathrm{min}}\ \underset{D}{\mathrm{max}} \ [\mathbb{E}_{\mathbf{y}\thicksim{p_{\mathbf{y}}{}}}(log D_\mathbf{y}) \\
 + \mathbb{E}_{\mathbf{\hat{y}}\thicksim{p_{\mathbf{\hat{y}}}{}}}(log (1-D_{\mathbf{\hat{y}}}))].
\end{split}
\label{0}
\end{equation}
Where CEL represent Cross-Entropy Loss and $\mathbb{E}$ is expectation. The model parameter sets $\Theta_{G}$ and $\Theta_{D}$ are updated during the process of minimization and maximization of the loss function. This competitive nature of learning is known as min-max game or zero-sum game. After a certain period of learning, probability distribution, ${p_\mathbf{\hat{y}}}$ becomes similar to ${p_\mathbf{y}}$ that resulted in an equilibrium state known as Nash-equilibrium state. In this situation, the discriminator provides a probability value of 0.5, which implies the generated synthetic samples have 50\% probability of being fake or real. Therefore, the generator network, discriminator network, and the loss-function plays an essential role in the GAN model to achieve an equilibrium state.
\par
In GAN models, it is nearly intractable to explicitly trace the evidence of probability distribution of generated and original data samples. Loss function provides an indirect estimation of the difference between generated and target data samples' underlying probability distribution. Hence, loss function has a significant impact on the study of GAN research to various problems. Recent studies reveal that the appropriate selection of loss function in GAN models is an emerging and promising research direction for generating high-quality synthetic data \cite{lOSS1,lOSSfunc2,loss-3}.

\subsection{Voice Conversion (VC) using GAN}
\begin{figure}[ht]
    \centering
    \includegraphics[height=0.27\linewidth, width=0.65\linewidth]{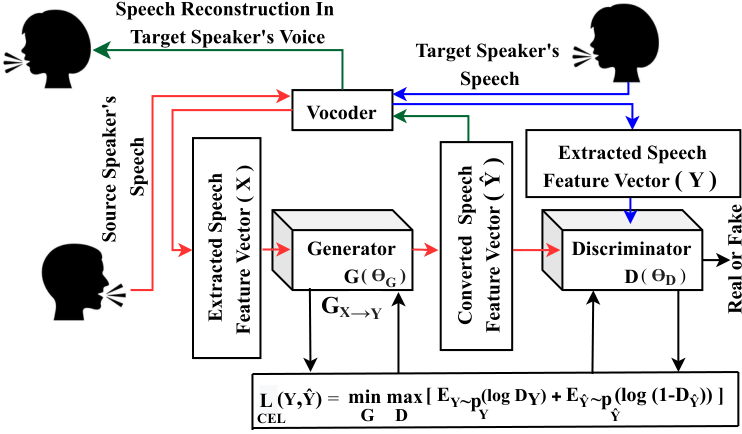} 
    \caption{Basic framework of a typical GAN based VC system}
    \label{GAN-VC1}
\end{figure}

Recent studies revealed that GAN models are the most promising one for VC task as compare to traditional models \cite{review-paper-1}. The basic framework for VC using a typical GAN model is shown in Fig. \ref{GAN-VC1}.
\par
In Fig. \ref{GAN-VC1}, vocoders are used for speech feature extraction from a source and a target speakers' speech data. The extracted speech features of the source speaker and the target speaker are then fed to the generator and the discriminator interms of speech feature vectors, respectively. The generator converts the source speaker's speech features $\mathbf{x}$ into a particular target speaker's speech features represented by $\mathbf{\hat{y}}$. On the other hand, the discriminator is trained with the target speaker's real speech features represented by $\mathbf{y}$. The discriminator distinguishes the converted speech features ($\mathbf{\hat{y}}$) from real speech features ($\mathbf{y}$) with the help of loss function. In general, Cross Entropy Loss (CEL) is used to train a GAN model. The parameters of the generator and the discriminator are updated during the training process through maximization and minimization of the loss function. After achieving equilibrium state of the GAN model, the vocoder reconstructs the audible speech from the converted speech features without changing the linguistic contents.
\subsection{Related Work} 
The proposed ALGAN-VC model mainly focused on one-to-one VC task. The related work includes the kinds of literature that rely on GAN-based approaches for one-to-one VC. Kaneko and Kameoka \cite{cyclegan-vc-1} implemented original CycleGAN \cite{CycleGAN2017} framework for non-parallel one-to-one VC task. They considered MCEP features for speech feature conversion and WORLD vocoder \cite{world} for the reconstruction of audible speech. The model is equipped with a residual connection-based 1D gated CNN for the generator network and a 2D gated CNN for the discriminator network and similar loss components as in the CycleGAN. But the model generates buzzy sounding speech in case of inter gender VC for over smoothing problem. Seshadri et al. \cite{cyclegan-vc-4} incorporated PML vocoder \cite{PML} as the first approach in CycleGAN-VC framework for both speech feature extraction and speech reconstruction. This resulted nominal improvement of the generated speech samples. The Improved version of CycleGAN-VC was proposed in CycleGAN-VC2 \cite{cyclegan-vc-2}. Here, a two step adversarial loss is used to deal with the over smoothing problem as observed in CycleGAN-VC. Moreover, an improved generator and a PatchGAN discriminator is incorporated to enhance the effectiveness of the model. Their model performed well for intra-gender VC with less naturalness. In CycleGAN-VC-GP \cite{cyclegan-vc-103}, zero-centered gradient penalties and combined fundamental frequency with the spectrum are considered to ensure the convergence of GAN model and improvement of the prosody conversion in CycleGAN-VC2 framework. In CycleGAN-VC3 \cite{cyclegan-vc-3}, mel-spectrogram conversion introduced instead of MCEP conversion as speech feature conversion. In this method, time-frequency adaptive normalization is used to adjust the scale and bias of the converted features. They validated their models performance on mono-lingual intra and inter gender VC. In MelGAN-VC \cite{deep-learning-8}, spectrogram conversion incorporated with siamese network. In the model, travel loss is combined with adversarial loss and identity loss for preserving speech information. They also used Griffin-Lim algorithm \cite{Griffin-Lim} for audible speech reconstruction from spectrograms. Du et. al \cite{cyclegan-vc-7} proposed Spectrum-Prosody-CycleGAN framework for cross-lingual VC. The model incorporated continuous wavelet transform (CWT) decomposition technique for $F_0$ modeling instead linear transformation of $F_0$. 
\par
The above mentioned studies are mainly focused on the effectiveness of architectural modifications of GAN and different speech features to improve speaker similarity and speech quality on mono-lingual speech data. However, there is a scope to explore other components of GAN models such as loss function, activation function, and learning rate for both cross-lingual and mono-lingual speech data.

\section{Proposed ALGAN-VC model}\label{proposed-approach}
\begin{figure}[ht]
    \centering
    \includegraphics[height=0.55\textwidth, width=0.7\textwidth]{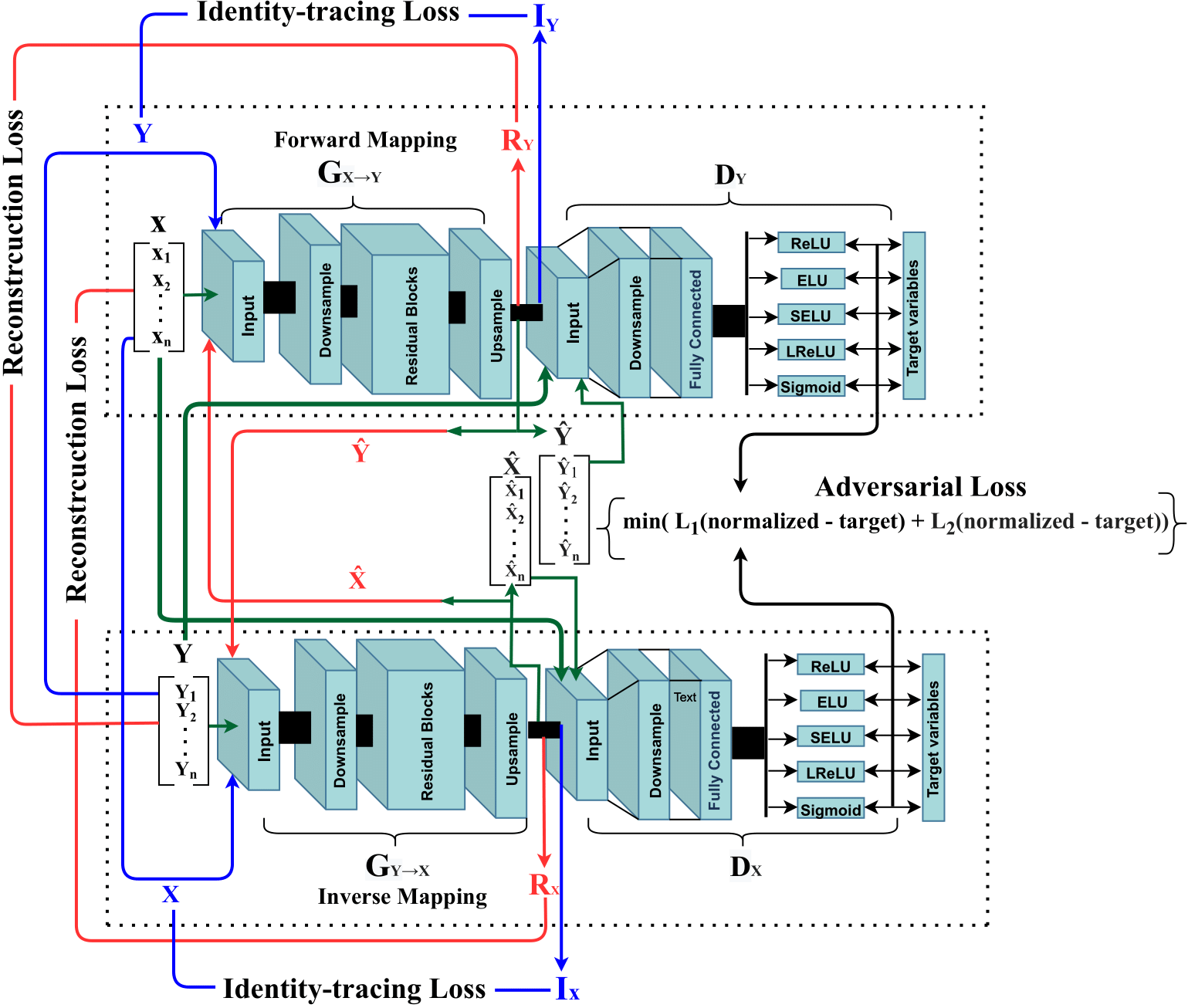}
    \caption{Schematic diagram of the proposed ALGAN-VC model. Here, identity of the loss functions are represented by uniquely coloured arrows.}
    \label{basic-frame-work}
\end{figure}

In this section, framework of the proposed ALGAN-VC model is described in detail. The proposed model consists of two generators $G_{\mathbf{x} \rightarrow \mathbf{y}}$ and $G_{\mathbf{y} \rightarrow \mathbf{x}}$, and two discriminators $D_{\mathbf{x}}$ and $D_{\mathbf{y}}$, where $\mathbf{x}$ and $\mathbf{y}$ represent speech feature vectors extracted from two different speakers speech sample. The overall process of ALGAN-VC depicted in Fig. \ref{basic-frame-work}.
\par
The proposed model learned both forward and inverse mapping simultaneously. In forward mapping ${\mathbf{x}\rightarrow \mathbf{y}}$, a generator ($G_{\mathbf{x} \rightarrow \mathbf{y}}$) served a transformation from a source speakers' speech features ${\mathbf{x}}$ to a particular target speaker's speech features $\mathbf{y}$. The mapping function $G_{\mathbf{x} \rightarrow \mathbf{y}}$ converted the source speech features $\mathbf{x}$ into speech features $\mathbf{\hat{y}}$ (i.e $G_{\mathbf{x}\rightarrow \mathbf{y}}(\mathbf{x}))$ in order to match the target speaker's speech features $\mathbf{y}$. Then, the discriminator $D_{\mathbf{y}}$ is trained with target speaker's speech features $\mathbf{y}$ and distinguished the converted speech features $\mathbf{\hat{y}}$ from $\mathbf{y}$. Similarly, in inverse mapping ${\mathbf{y}\rightarrow\mathbf{x}}$, a generator (${G_{\mathbf{y}\rightarrow \mathbf{x}}}$ served a transformation from source speaker's speech features ${\mathbf{y}}$ to target speaker's speech features $\mathbf{x}$. The mapping function ${G_{\mathbf{y}\rightarrow \mathbf{x}}}$ converted the speech features $\mathbf{y}$ into speech features $\mathbf{\hat{x}}$ (i.e ${G_{\mathbf{y}\rightarrow \mathbf{x}}(\mathbf{y}))}$ inorder to match the target speaker's speech features $\mathbf{x}$. Then, the discriminator ${D_{{\mathbf{x}}}}$ is trained with target speaker's speech features $\mathbf{x}$ and distinguished the converted speech features $\mathbf{\hat{x}}$ from $\mathbf{x}$.
\par
Moreover, the proposed model includes three main components where we make some contributions to enhance speech feature learning ability for efficient and effective voice conversion. Firstly, a dense residual network (DRN) like architecture is proposed in the generator network of the ALGAN-VC model. This architecture improves residual speech feature mapping ability from source speaker to target speaker. Secondly, the total loss of our model is obtained as the summation of three losses such as adversarial loss, reconstruction loss, and identity-tracing loss. An adaptive learning mechanism is proposed for computing adversarial loss, a linear combination of ${L_1}$ and ${L_2}$ loss functions. Reconstruction loss is calculated using ${L_1}$ loss between reconstructed speech features ${\mathcal{R}_{\mathbf{x}}}$ (i.e ${G_{\mathbf{y}\rightarrow \mathbf{x}}(G_{\mathbf{x}\rightarrow \mathbf{y}}(\mathbf{x}))}$) and speech feature vector $\mathbf{x}$, and ${\mathcal{R}_{\mathbf{y}}}$ (i.e ${G_{\mathbf{x}\rightarrow \mathbf{y}}(G_{\mathbf{y}\rightarrow \mathbf{x}}(\mathbf{y}))}$) and $\mathbf{y}$. Furthermore, identity-tracing loss is also calculated using ${L_1}$ loss between identity-traced ${\mathcal{I}_{\mathbf{x}}}$ (i.e ${G_{\mathbf{y}\rightarrow \mathbf{x}}({\mathbf{x}})}$) and $\mathbf{x}$, and ${\mathcal{I}_{\mathbf{y}}}$ (i.e ${G_{\mathbf{x}\rightarrow  \mathbf{y}}({\mathbf{y}})}$) and $\mathbf{y}$, respectively. Thirdly, a boosted learning rate strategy is proposed for updating the learning rate of the generator and the discriminator to enhances the feature learning ability of the proposed model.
\par
The Details of DRN like architecture, adaptive learning approach and boosted learning rate strategy for the proposed ALGAN-VC model are described below.

\subsection{Dense Residual network architecture}
In residual connection based CNN models (ResNets) \cite{Resnet}, layers outputs are aggregated via a summation operation and feed-forward to the succeeding layers as inputs for enhancing the performance of the models. It has the potential ability to improve the gradient flow and prevent the model from vanishing gradient and exploding gradient problems due to the presence of skip connections as residual links while backpropagation is performed \cite{Resnets-for-speech-enhancements}, \cite{Resnets-for-speech-enhancements2}. Whereas, densely connected CNN (DenseNet) \cite{DenseNet-new} model consists of the dense aggregation of layers that provide direct feature re-usage as deeper layers can access the outputs of shallower layers. Therefore, merging the advantages of both the ResNets and DenseNets can improve feature learning ability and the whole training process of a model. With these benefits, a Dense Residual Network (DRN) like architecture is incorporated into the generator network in our proposed model.
\par
\begin{figure*}[ht]
    \centering
    \includegraphics[height=7.5cm, width=17.5cm]{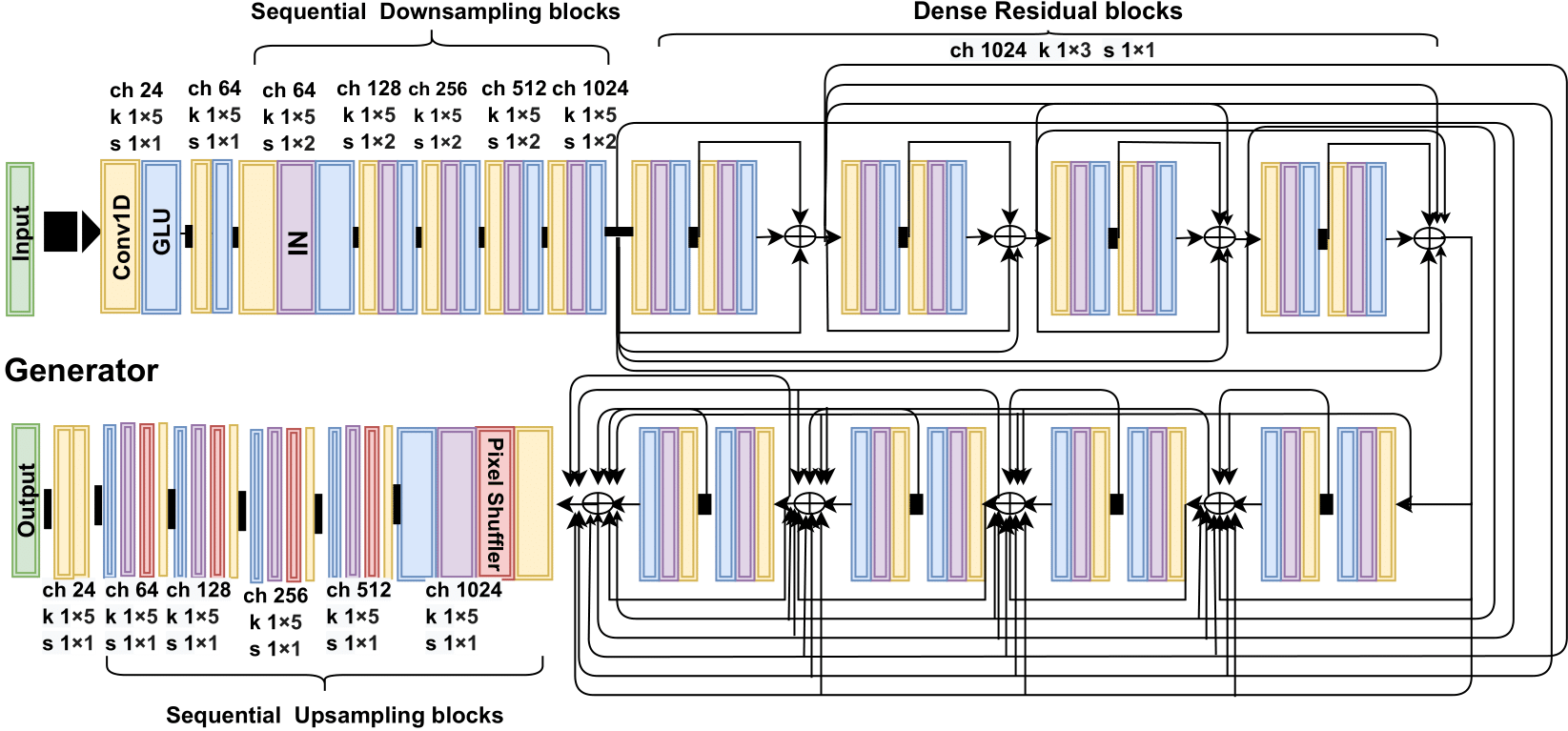}
    \caption{ Dense Residual Network (DRN) based generator network architecture of the ALGAN-VC model. In each blocks ch, k, s and IN represents the number of channels, kernel size , stride and instance normalization respectively. Each unique color code represents respective blocks' identity.}
    \label{generator}
\end{figure*}

The generator network of the proposed ALGAN-VC model is consists of three blocks such as sequential downsampling blocks, dense residual blocks and sequential upsampling blocks as shown in  Fig. \ref{generator}.
\subsubsection{Sequential Downsampling Blocks}
Sequential downsampling blocks are consists of several convolution units termed as downsampling block. It can be expressed as composite function, ${h^{d}(\cdot)}$ ($d$ represent downsampling block) of three operations: 1D convolution ($Conv$), instance normalization ($IN$), and gated linear unit ($GLU$). Hence, the \textit{i-th} downsampling block can be expressed mathematically as
\begin{equation}
 {h}_{i}^{d}(\mathbf{x})=IN(GLU(Conv(\mathbf{x}))),
\label{002}
\end{equation}
where $\mathbf{x}$ represent the input feature vector. 
\subsubsection{Dense Residual Blocks}
The output of the last sequential downsampling block flows to the dense residual blocks in DRN architecture as input. Moreover, the proposed DRN architecture consists of several dense residual blocks. So, each dense residual block can represent a composite function, ${H^{r}(\cdot)}$ ($r$ for residual block) of two convolution units (i.e ${{h}^{r}}$ and ${{h}^{{r}^{'}}}$) and mathematically expressed as
\begin{equation}
\begin{gathered}
 {H}^{r}_{j}(\mathbf{x})={h}^{{r}^{'}}_{j}({h}^{r}_{j}(\mathbf{x})),
\label{01}
\end{gathered}
\end{equation}
where ${H}^{r}_{j}$ represent the \textit{j-th} dense residual block. ${{h}_{j}^{r}}$ and ${{h}^{{r}^{'}}_{j}}$ represent the first and second convolution unit of the ${j}^{th}$ block. Additionally, each convolution unit of dense residual block performs the three operations the same as in sequential downsampling block. Therefore, the convolution unit, ${h^{r}(\cdot)}$ of the \textit{j-th} block over input feature vector $\mathbf{x}$ can be formulated as
\begin{equation}
\begin{gathered}
 {h}^{r}_{j}(\mathbf{x})=IN(GLU(Conv(\mathbf{x}))).
\label{02}
\end{gathered}
\end{equation}
Therefore, the output of the \textit{j-th} residual block is defined as 
\begin{equation}
\begin{gathered}
 {O}_{j}={H}^{r}_{j}(\mathbf{x}),
\label{03}
\end{gathered}
\end{equation}
\par
Now, the output of the \textit{1-st} block can be written explicitly as
\begin{equation}
\begin{gathered}
 {O}_{1}={H}^{r}_{1}(\mathbf{x}), \quad \text{where} \quad {H}^{r}_{1}={h}^{{r}^{'}}_{1}({h}^{r}_{1}(\mathbf{x})).
\label{04}
\end{gathered}
\end{equation}
The ${{O}_{1}}$ is added with the first convolution unit output, ${{h}^{r}_{1}}$ and skip connection $\mathbf{x}$ and collectively treated as an input to $2$\textit{-nd} residual block to produce output $O_2$. Therefore, the output $O_2$ is expressed as
\begin{equation}
\begin{gathered}
{O}_{2}={H}^{r}_{2}({O}_{1}+{h}^{r}_{1}(\mathbf{x})+\mathbf{x}) \&
{O}_{2}={H}^{r}_{2}({I}_{1}) \&
{I}_{1}={O}_{1}+{h}^{r}_{1}(\mathbf{x})+\mathbf{x} \quad \text{and} \quad {H}^{r}_{2}={h}^{{r}^{'}}_{2}({h}^{r}_{2}({I}_{1})).
\label{05}
\end{gathered}
\end{equation}
Here, ${{I}_{1}}$ represent the $1$\textit{-st} residual summation. It indicates that the input and the output of a particular dense residual block of the DRN architecture depend on all preceding block's output. Therefore, output of the $n$\textit{-th} residual block can be expressed recursively as
\begin{equation}
\begin{gathered}
 {O}_{n}={H}^{r}_{n}({I}_{n-1}) \quad \text{and} \quad {H}^{r}_{n}={h}^{{r}^{'}}_{n}({h}^{r}_{n}({I}_{n-1})),\\
 {I}_{n-1}={O}_{n-1}+{h}^{r}_{n-1}({I}_{n-1})+\sum_{s=1}^{n-2} {I}_{s}+\mathbf{x}.
\label{08}
\end{gathered}
\end{equation}
The residual summation (${I_{n-1}}$) exhibits cumulative feature flow, enhancing the feature learning ability of the generator network in the proposed ALGAN-VC model.
\subsubsection{Sequential Upsampling Blocks}
The output of the final residual summation is feed to sequential upsampling blocks as input. Each of the block represent a composite function, ${h^{u}(\cdot)}$ ($u$ for upsampling) of four operations such as 1D convolution ($Conv$), pixel shuffler ($PS$), instance normalization ($IN$), and gated linear unit ($GLU$). So, the ${k}$\textit{-th} downsampling block for the input feature vector $\mathbf{x}$ is expressed mathematically as
\begin{equation}
\begin{gathered}
 {h}_{k}^{u}(\mathbf{x})=IN(PS(GLU(Conv(\mathbf{x})))).
\label{99}
\end{gathered}
\end{equation}
\par
The use of downsampling and upsampling blocks in the proposed model helps us to make the model computationally less expensive and provides high-quality style transfer from source to target speech features.
\subsection{Adaptive learning approach}
An adaptive learning approach is proposed for the efficient training of our model. It aims to compute the adversarial loss adaptively. This section mainly focuses on the computation of the adversarial loss function in the ALGAN-VC model. Furthermore, reconstruction loss and identity-tracing loss are also discussed, respectively.  \subsubsection{Adversarial loss}
\begin{figure*}[ht]
    \centering
    \includegraphics[height=5.25cm, width=17cm]{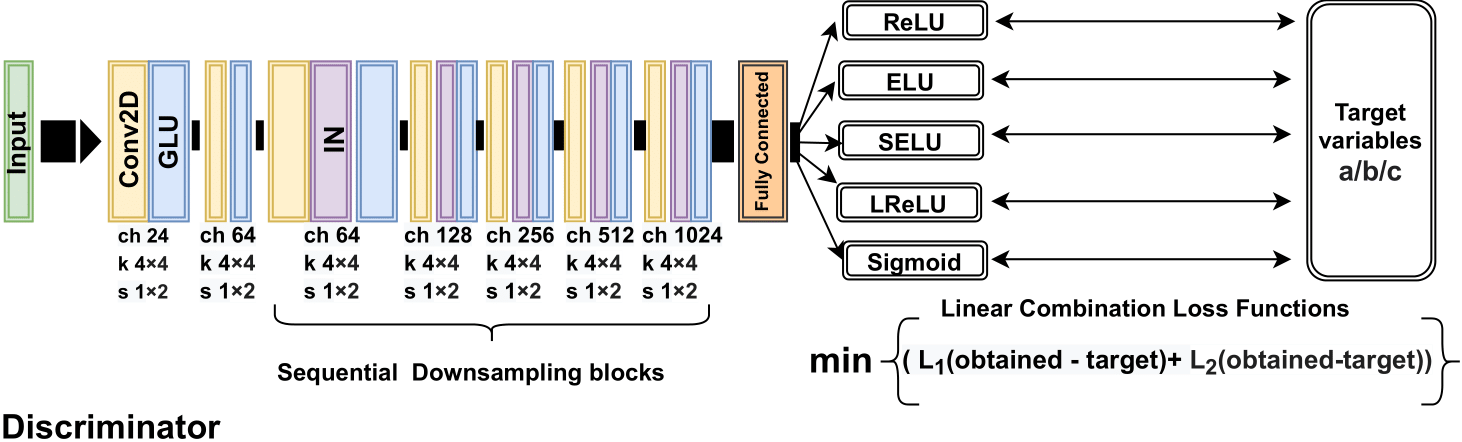}
    \caption{Discriminator network of the ALGAN-VC model. In each blocks ch, k, s and IN represents the number of channels, kernel size , stride and instance normalization respectively. Each unique color code represents respective blocks' identity.}
    \label{discriminator}
\end{figure*}

The adversarial loss is formulated as a linear combination of $L_1$ and $L_2$ loss, computed using a normalization technique based on multiple activation functions. It is well known that different loss functions provide unlike underlying distributions in GAN models \cite{GAN-LOSS-FUNCTION}. To owing these benefits, a linear combination of loss functions is considered in our study for better aspect of tracing underlying distribution of input data. Both $L_1$ and $L_2$ losses are calculated based on the difference between the normalized output and the respective target label. The normalized outputs are obtained by using multiple activation functions \cite{Adaptive-activation-function-1} from the output layer of the discriminator. The target labels are derived mathematically with probability divergence theorem (discussed latter in this section). The $L_1$ and $L_2$ losses have been shown their potential ability to improved the perceptual quality of speech and well performed on the modification of speaker individuality and generates transformed voice with high speech quality in case of speech enhancement \cite{Sppech-L1,Speech-L1-1,L2}. For these advantages, the adversarial loss is designed in terms of a linear combination of the $L_1$ and $L_2$ loss to achieve better speech perceptual quality and audio quality in the proposed model for voice conversion. The adversarial loss computation is depicted in Fig. \ref{discriminator}.
\par
The adversarial loss of the generator or the discriminator is expressed as
\begin{equation}
\begin{gathered}
 \mathcal{L}=\alpha\times {L_1} + \beta\times {L_2}, \quad {\alpha + \beta} =1.
\label{999}
\end{gathered}
\end{equation}
Where, $\alpha$ and $\beta$ are two constants. The minimization of adversarial loss maximizes the similarity between normalized outputs and target labels. For the simplicity, the output of the discriminators and their corresponding target label are denoted as ${D_{\mathbf{x}}(\mathbf{x})}$, ${D_{\mathbf{y}}(\mathbf{y})}$ and $b$ in case of real speech data samples $\mathbf{x}$ and $\mathbf{y}$. Similarly, for the converted (fake) speech data samples $\mathbf{\hat{x}}$ and $\mathbf{\hat{y}}$, the output of the discriminators and their corresponding target label are denoted as ${D_{\mathbf{x}}(\mathbf{\hat{x}})}$, ${D_{\mathbf{y}}(\mathbf{\hat{y}})}$ and $a$. Therefore, the adversarial loss, ${{\mathcal L}_{{D}_{\mathbf{ x}}}}$ and ${{\mathcal L}_{{D}_{\mathbf y}}}$ for the discriminator, ${D_{\mathbf{y}}(\mathbf{x})}$ and ${D_{\mathbf{y}}(\mathbf{\textit{y}})}$ are formulated as 
\begin{flalign}
 &{\mathcal{L}_{D_{\mathbf{x}}}} = \frac{1}{2}{\mathbb{E}_{\mathbf{x}\sim {p_{\mathbf{x}}}(\mathbf{x})}}{[\alpha L_1+\beta L_2]} {(\text{for real data})}&\nonumber\\&+\frac{1}{2} {\mathbb{E}_{\mathbf{y}\sim{p_{\mathbf{y}}}(\mathbf{y})}}{[\alpha L_1 + \beta L_2]}{(\text{for fake data})}.&\nonumber\\
&{\mathcal{L}_{D_{\mathbf{x}}}}=\frac{1}{2}{\mathbb{E}_{\mathbf{x}\sim{p_{\mathbf{x}}}(\mathbf{x})}}[\alpha||({D_{\mathbf{x}}}(\mathbf{x}))-{b}|{|_1}+\beta||({D_{\mathbf{x}}}(\mathbf{x}))-{b}|{|_2}]&\nonumber\\&+\frac{1}{2}{\mathbb{E}_{\mathbf{y}\sim{p_{\mathbf{y}}}(\mathbf{y})}}[\alpha||({D_{\mathbf{x}}}(\mathbf{\hat{x}}))-{a}|{|_1}+\beta||({D_{\mathbf{x}}}(\mathbf{\hat{x}}))-{a}|{|_2}]&\nonumber\\
 & = \frac{1}{2}{\mathbb{E}_{\mathbf{x}\sim{p_{\mathbf{x}}}(\mathbf{x})}}[\alpha||({D_{\mathbf{x}}}(\mathbf{x}))-{b}|{|_1}+\beta||({D_{\mathbf{x}}}(\mathbf{x}))-{b}|{|_2}]&\nonumber\\&+\frac{1}{2}{\mathbb{E}_{\textbf{y}\sim {p_{\textbf{y}}}(\textbf{y} )}}[\alpha||({D_{\mathbf{x}}}(G_{\mathbf{y}\rightarrow\mathbf{x}}({\mathbf{y}})))-{a} |{|_1}&\nonumber\\&+\beta||({D_{\mathbf{x}}}(G_{\mathbf{y} \rightarrow \mathbf{x}}({\mathbf{y}})))-{a}|{|_2}],
\label{eq:123}  
\end{flalign}
\begin{flalign}
&{\mathcal{L}_{D_{\mathbf{y}}}}=\frac{1}{2}{\mathbb{E}_{\mathbf{y}\sim{p_{\mathbf{y}}}(\mathbf{y})}}[\alpha||({D_{\mathbf{y}}}(\mathbf{y}))-{b}|{|_1}+\beta||({D_{\mathbf{y}}}(\mathbf{y}))-{b}|{|_2}]&\nonumber\\&+\frac{1}{2}{\mathbb{E}_{\mathbf{y}\sim{p_{\mathbf{y}}}(\mathbf{y})}}[\alpha||({D_{\mathbf{y}}}(\mathbf{\hat{y}}))-{a}|{|_1}+\beta||({D_{\mathbf{y}}}(\mathbf{\hat{y}}))-{a}|{|_2}]&\nonumber\\
 & = \frac{1}{2}{\mathbb{E}_{\mathbf{y}\sim{p_{\mathbf{y}}}(\mathbf{y})}}[\alpha||({D_{\mathbf{y}}}(\mathbf{y}))-{b}|{|_1}+\beta||({D_{\mathbf{y}}}(\mathbf{y}))-{b}|{|_2}]&\nonumber\\&+\frac{1}{2}{\mathbb{E}_{\textbf{x}\sim {p_{\textbf{x}}}(\textbf{x})}}[\alpha||({D_{\mathbf{y}}}(G_{\mathbf{x}\rightarrow\mathbf{y}}({\mathbf{x}})))-{a}|{|_1}&\nonumber\\&+\beta||({D_{\mathbf{y}}}(G_{\mathbf{x}\rightarrow \mathbf{y}}({\mathbf{x}})))-{a}|{|_2}].
\label{eq:1231}  
\end{flalign}
Similarly, the adversarial loss for the generators, $G_{\mathbf{x}\rightarrow\mathbf{y}}$ and $G_{\mathbf{y}\rightarrow\mathbf{x}}$ can be formulated as
\begin{flalign}
 &{\mathcal{L}_{G_{\mathbf{x}\rightarrow\mathbf{y}}}}=\frac{1}{2}
  {\mathbb{E}_{\textbf{y}\sim{p_{\mathbf{y}}}(\mathbf{y})}}[\alpha||({D_{\mathbf{y}   }}(\mathbf{y}))-{{c}}|{|_1}+\beta||({D_{\mathbf{y}}}(\mathbf{y}))-{{c}}|{|_2}]&\nonumber\\&+\frac{1}{2}{\mathbb{E}_{\mathbf{x}  \sim {p_{\mathbf{x}}}(\mathbf{x} )}}[\alpha||({D_{\mathbf{y}   }}(G_{\mathbf{x}\rightarrow\mathbf{y}}(\mathbf{x})))-{{c}}|{|_1}&\nonumber\\&+ \beta||({D_{\mathbf{y}}}(G_{\mathbf{x}\rightarrow\mathbf{y}}(\mathbf{x})))-{{c}}|{|_2}],
 \label{eq:1255}  
\end{flalign}
\begin{flalign}
 &{\mathcal{L}_{G_{\mathbf{y}\rightarrow\mathbf{x}}}}=\frac{1}{2}
  {\mathbb{E}_{\textbf{x}\sim{p_{\mathbf{x}}}(\mathbf{x})}}[\alpha||({D_{\mathbf{x}   }}(\mathbf{x}))-{{c}}|{|_1}+\beta||({D_{\mathbf{x}}}(\mathbf{x}))-{{c}}|{|_2}]&\nonumber\\&+\frac{1}{2}{\mathbb{E}_{\mathbf{y} \sim {p_{\mathbf{y}}}(\mathbf{y})}}[\alpha||({D_{\mathbf{x}   }}(G_{\mathbf{y}\rightarrow\mathbf{x}}(\mathbf{y})))-{{c}}|{|_1}&\nonumber\\&+ \beta||({D_{\mathbf{x}}}(G_{\mathbf{y}\rightarrow\mathbf{x}}(\mathbf{y})))-{{c}}|{|_2}],
 \label{eq:1245}  
\end{flalign}
where $c$ is the target data label.
\par
The values of $a$, $b$, and $c$ can be determined by using either forward or inverse mapping. In the forward mapping ($\mathbf{x}\rightarrow\mathbf{y}$), the optimal discriminator, ${D}^{*}_{\mathbf y}$ is determined by taking the derivative of Eq. (\ref{eq:1231}) with respect to ${D}_{\mathbf{y}}$ and $\frac{d{\mathcal L}_{{D}_{\mathbf y}}}{d D_{\mathbf y}}=0$ keeping the fixed generator ${G_{\mathbf{x}\rightarrow \mathbf{y}}}$. Therefore, the optimal discriminator, ${D}^{*}_{\mathbf y}$ is
\begin{equation}
\begin{gathered}
   {\mathcal D^{*}_{\mathbf y} ({\mathbf y})} = \frac{(\beta {b}-0.5\alpha)p_{\mathbf{y}}(\mathbf{y})+(\beta {a}-0.5\alpha)p_{\mathbf{\hat{y}}}(\mathbf{y})}{\beta(p_{\mathbf{y}}(\mathbf{y})+p_{\mathbf{\hat{y}}}(\mathbf{y}))}.
\end{gathered} 
\label{eq:71}
\end{equation}
After putting ${D}^{*}_{\mathbf y}$ in Eq. (\ref{eq:1255}), it is reformulated as 
\begin{flalign}
  & {\mathcal{L}_{G_{\mathbf{x}\rightarrow\mathbf{y}}}} = \frac{1}{2}
  {\mathbb{E}_{\mathbf{y}\sim {p_{\mathbf{y}}}}}[\alpha||({D^{*}_{\mathbf{y}}}(\mathbf{y}))- {c}|{|_1} +\beta ||({D^{*}_{\mathbf{y}}}(\mathbf{y}))-{c}|{|_2}] & \nonumber \\[7pt]
  &+\frac{1}{2} {\mathbb{E}_{\mathbf{y}\sim {p_{\mathbf{\hat{y}}}}}}[\alpha ||({D_{\mathbf{y}   }}^{*}(\mathbf{y}))-{c}|{|_1} +\beta ||({D^{*}_{\mathbf{y}}}(\mathbf{y})-{c} |{|_2}],& \nonumber  \\[7pt]
  & {\mathcal{L}_{G_{\mathbf{x}\rightarrow\mathbf{y}}}} = \frac{1}{2}
  {\mathbb{E}_{\mathbf{y}\sim{p_{\textbf{y}}}}}
  [\alpha(\frac{{b}p_{\mathbf{y}}(\mathbf{y})+{a}p_{\mathbf{\hat{y}}}(\mathbf{y})}{p_{\mathbf{y}}(\mathbf{y})+p_{\mathbf{\hat{y}}}(\mathbf{y})}) - {{c}}) &\nonumber  \\[7pt] & + \beta ((\frac{{b}p_{\mathbf{y}}(\mathbf{y})+{a}p_{\mathbf{\hat{y}}}(\mathbf{y})}{p_{\mathbf{y}}(\mathbf{y})+p_{\mathbf{\hat{y}}}(\mathbf{y})}) - {{c}})^{2}] &\nonumber  \\[7pt] &+\frac{1}{2}
  {\mathbb{E}_{\mathbf{y}  \sim {p_{\mathbf{\hat{y}}}}}}
  [\alpha((\frac{{b}p_{\mathbf{y}}(\mathbf{y})+{a}p_{\mathbf{\hat{y}}}(\mathbf{y})}{p_{\mathbf{y}}(\mathbf{y})+p_{\mathbf{\hat{y}}}(\mathbf{y})}) - {{c}}) &\nonumber \\[7pt] & + \beta ( (\frac{{b}p_{\mathbf{y}}(\mathbf{y})+{a}p_{\mathbf{\hat{y}}}(\mathbf{y})}{p_{\mathbf{y}}(\mathbf{y})+p_{\mathbf{\hat{y}}}(\mathbf{y})}) - {{c}})^{2}].  
\label{eq:73}  
\end{flalign}
Eq. (\ref{eq:73}) is obtained by ignoring the constants associated in Eq. (\ref{eq:71}). Therefore, minimization of the adversarial loss ($\mathcal{L}_{G_{\mathbf{x}\rightarrow\mathbf{y}}}$) is defined as
\begin{flalign}
  & \min_{{G_{\mathbf{x}\rightarrow\mathbf{y}}}}{\mathcal{L}_{G_{\mathbf{x}\rightarrow\mathbf{Y}}}} =  \frac{1}{2}
  \int_{\mathbf{y}}p_{\mathbf{y}}(\mathbf{y})
  [\alpha(\frac{{b}p_{\mathbf{y}}(\mathbf{y})+{a}p_{\mathbf{\hat{y}}}(\mathbf{y})}{p_{\mathbf{y}}(\mathbf{y})+p_{\mathbf{\hat{y}}}(\mathbf{y})}) - {{c}})  &\nonumber \\[7pt] & + \beta ((\frac{{b}p_{\mathbf{y}}(\mathbf{y})+{a}p_{\mathbf{\hat{y}}}(\mathbf{y})}{p_{\mathbf{y}}(\mathbf{y})+p_{\mathbf{\hat{y}}}(\mathbf{y})}) - {{c}})^{2}]d\mathbf{y} &\nonumber \\[7pt] &+\frac{1}{2}\int_{\mathbf{y}}p_{\mathbf{\hat{y}}}(\mathbf{y})
  [\alpha((\frac{{b}p_{\mathbf{y}}(\mathbf{y})+{a}p_{\mathbf{\hat{y}}}(\mathbf{y})}{p_{\mathbf{y}}(\mathbf{y})+p_{\mathbf{\hat{y}}}(\mathbf{y})}) - {{c}}) &\nonumber \\[7pt] & + \beta ( (\frac{{b}p_{\mathbf{y}}(\mathbf{y})+{a}p_{\mathbf{\hat{y}}}(\mathbf{y})}{p_{\mathbf{y}}(\mathbf{y})+p_{\mathbf{\hat{y}}}(\mathbf{y})}) - {{c}})^{2}]d\mathbf{y},&\nonumber \\[7pt]
 &= \int_{\mathbf{y}}p_{\mathbf{y}}(\mathbf{y})
  [\alpha(\frac{{b}p_{\mathbf{y}}(\mathbf{y})+{a}p_{\mathbf{\hat{y}}}(\mathbf{y})}{p_{\mathbf{y}}(\mathbf{y})+p_{\mathbf{\hat{y}}}(\mathbf{y})}) - {{c}}) &\nonumber \\[7pt] & + \beta ((\frac{{b}p_{\mathbf{y}}(\mathbf{y})+{a}p_{\mathbf{\hat{y}}}(\mathbf{y})}{p_{\mathbf{y}}(\mathbf{y})+p_{\mathbf{\hat{y}}}(\mathbf{y})}) - {{c}})^{2}]d\mathbf{y} &\nonumber \\[7pt] &+\frac{1}{2}\int_{\mathbf{y}}p_{\mathbf{\hat{y}}}(\mathbf{y})
  [\alpha((\frac{{b}p_{\mathbf{y}}(\mathbf{y})+{a}p_{\mathbf{\hat{y}}}(\mathbf{y})}{p_{\mathbf{y}}(\mathbf{y})+p_{\mathbf{\hat{y}}}(\mathbf{y})}) - {{c}}) &\nonumber \\[7pt] & + \beta ( (\frac{{b}p_{\mathbf{y}}(\mathbf{y})+{a}p_{\mathbf{\hat{y}}}(\mathbf{y})}{p_{\mathbf{y}}(\mathbf{y})+p_{\mathbf{\hat{y}}}(\mathbf{y})}) - {{c}})^{2}]d\mathbf{y}.
\label{eq:744}  
\end{flalign}

After solving Eq. (\ref{eq:744}) by ignoring the constants ${\alpha}$ and ${\beta}$, we get
\begin{flalign}
  &\min_{{G_{\mathbf{x}\rightarrow\mathbf{y}}}}{\mathcal{L}_{G_{\mathbf{x}\rightarrow\mathbf{Y}}}} = &\nonumber \\
  &\frac{1}{2}
  [\int_{\mathbf{y}}(b-c)(p_{\mathbf{y}}(\mathbf{y})+p_{\mathbf{\hat{y}}}(\mathbf{y}))-p_{\mathbf{\hat{y}}}(\mathbf{y})(b-a)d\mathbf{y}\hfill &\nonumber \\
    &+\int_{\mathbf{y}}\frac{\{(b-c)(p_{\mathbf{y}}(\mathbf{y})+p_{\mathbf{\hat{y}}}(\mathbf{y}))-p_{\mathbf{\hat{y}}}(\mathbf{y})(b-a)\}^2}{p_{\mathbf{y}}(\mathbf{y})+p_{\mathbf{\hat{y}}}(\mathbf{y})}d\mathbf{y}].
\end{flalign}
Putting, ${(b-c)=1}$ and ${(b-a)=2}$ and therefore, the above equation is written as
 \begin{flalign}
  &\min_{{G_{\mathbf{x}\rightarrow\mathbf{y}}}}{\mathcal{L}_{G_{\mathbf{x}\rightarrow\mathbf{Y}}}}= \frac{1}{2} \int_{\mathbf{y}}{(p_{\mathbf{y}}(\mathbf{y})+p_{\mathbf{\hat{y}}}(\mathbf{y})-2p_{\mathbf{\hat{y}}}(\mathbf{y})})d\mathbf{y} &\nonumber \\ &+\frac{1}{2} \int_{\alpha} 
 {(\frac{(2p_{\mathbf{\hat{y}}}(\mathbf{y})-(p_{\mathbf{y}}(\mathbf{y})+p_{\mathbf{\hat{y}}}(\mathbf{y})))^{2}}{(p_{\mathbf{y}}(\mathbf{y})+p_{\mathbf{\hat{y}}}(\mathbf{y}))}})d\mathbf{y} &\nonumber \\
  &= \frac{1}{2} \{ \int_{\alpha} 
  {(p_{\mathbf{y}}(\mathbf{y})-p_{\mathbf{\hat{y}}}(\mathbf{y})})d\mathbf{y} \} +{\chi}^{2}_{pearson} (p_{\mathbf{y}}+p_{\mathbf{\hat{y}}} \|\ 2p_{\mathbf{\hat{y}}}) &\nonumber \\
  &=\frac{1}{2} \{ \int_{\mathbf{y}} 
  {p_{\mathbf{y}}(\mathbf{y})}d\mathbf{y} - \int_{\mathbf{y}} 
  {p_{\mathbf{\hat{y}}}(\mathbf{y})}d\mathbf{y} \} &\nonumber \\ &+{\chi}^{2}_{pearson} (p_{\mathbf{y}}+p_{\mathbf{\hat{y}}} \|\ 2p_{\mathbf{\hat{y}}})  &\nonumber \\
  &=\frac{1}{2} \{ P_{\mathbf{y}}[\mathbf{y}]-P_{\mathbf{\hat{y}}}[\mathbf{y}] \} +{\chi}^{2}_{pearson} (p_{\mathbf{y}}+p_{\mathbf{\hat{y}}} \|\ 2p_{\mathbf{\hat{y}}}).
  \label{eq:100}
\end{flalign}
Usually, ${{0}-{1}}$ encoding scheme has been used for the target labels in classification algorithms. So, we used this encoding scheme in this paper to assign the target label values. Therefore, the equations ${b-c=1}$ and ${b-a=2}$ must satisfy the ${0-1}$ encoding scheme and the target label values are determined as ${b=1}$, ${c=0}$ and ${a-1}$. Finally, minimization of the adversarial loss ($\mathcal{L}_{G_{\mathbf{x}\rightarrow\mathbf{y}}}$) for the generator using forward mapping is expressed as two terms. First term corresponds to the difference between the probability density function of original samples and generated samples. The Chi-Square Pearson (${\chi}^{2}_{pearson}$) divergence between ($p_{\mathbf{y}}+p_{\mathbf{\hat{y}}}$) and $2p_{\mathbf{\hat{y}}}$ corresponds by the second term.
\par
The above adversarial losses are computed based on a particular activation function in the output layer of the discriminator. It is to be noted that we have used multiple activation functions in the output layer of the discriminator network to obtain the normalized output. Hence, the discriminator outputs ${D_{\mathbf{x}}}(\mathbf{x})$, ${{D}_{\mathbf{x}}(G_{\mathbf{y}\rightarrow\mathbf{x}}}(\mathbf{y}))$, ${D_{\mathbf{y}}}(\mathbf{y})$, and ${{D}_{\mathbf{y}}(G_{\mathbf{x}\rightarrow\mathbf{y}}}(\mathbf{x}))$ (shown in Eqs. (\ref{eq:123}, \ref{eq:1231}, \ref{eq:1255}, and \ref{eq:1245})) are normalized using multiple activation functions such as \textit{ReLU}, \textit{ELU}, \textit{SELU}, \textit{LReLU}, \textit{Sigmoid} in parallel and taking their minimum for computing the $L_1$ and $L_2$ loss. This computation procedure is depicted in Fig. \ref{discriminator}. The mentioned activation functions are considered in our study because of their non-linear transformation nature and broad applications perspective \cite{Activation-Function-Ensambling}. Therefore, the adversarial loss of the generator or the discriminator is formalized mathematically using the multiple activation functions as
\begin{equation}
\begin{array}{l@{}l}
\mathcal{L} = min{{}}[\alpha {||a(f)-O_t||}_1+\beta{||a(f)-O_t||}_2].
\end{array} 
\label{eq:5}
\end{equation}
Where, ${a(\cdot)\in \{\textit{ReLU}, \textit{ELU}, \textit{SELU}, \textit{LReLU}, \textit{Sigmoid}\}}$ is the activation function, ${f\in \{{D_{\mathbf{x}}}(\mathbf{x})}$, ${{D}_{\mathbf{x}}(G_{\mathbf{y}\rightarrow\mathbf{x}}}(\mathbf{y}))$, ${D_{\mathbf{y}}}(\mathbf{y})$, ${D}_{\mathbf{y}}(G_{\mathbf{x}\rightarrow\mathbf{y}}(\mathbf{x}))\}$ is the discriminator output and $O_t\in \{a, b, c\}$ is the target label output.
\par
Therefore, the adversarial loss for forward mapping is written as
\begin{equation}
\mathcal{L}_{\mathbf{x}\rightarrow\mathbf{y}} =  {\mathcal{L}_{\mathbf{G}_{\mathbf{x}\rightarrow\mathbf{y}}}}+{\mathcal{L}_{\mathbf{D}_{\mathbf{y}}}}.
\end{equation}
Similarly, the adversarial loss for inverse mapping is written as
\begin{equation}
{\mathcal{L}_{\mathbf{y}\rightarrow\mathbf{x}}} =  {\mathcal{L}_{\mathbf{G}_{\mathbf{y}\rightarrow\mathbf{x}}}}+{\mathcal{L}_{\mathbf{D}_{\mathbf{x}}}}.
\end{equation}
Thus, the total adversarial loss for ALGAN-VC model is
\begin{equation}
{\mathcal{L}_{\mathbf{Total}}} =  {\mathcal{L}_{\mathbf{x}\rightarrow\mathbf{y}}}+{\mathcal{L}_{\mathbf{y}\rightarrow\mathbf{x}}}
\end{equation}

\subsubsection{Reconstruction loss}
Despite the similarity measure between the two distributions, it is also necessary to preserve the contextual information of speech data during voice conversion task. For this purpose, reconstruction loss is considered in our model with the form of ${L}_{1}$ loss function. The reconstruction losses, ${\mathcal{{R}_{\text{rec}(\mathbf{x})}}}$ and ${\mathcal{{R}_{\mathbf{rec(y)}}}}$ of speech samples $\mathbf{x}$ and $\mathbf{y}$ are defined as
\begin{equation}
\begin{array}{l@{}l}
{\mathcal{{R}_{\mathbf{rec(x)}}}} &{}= {\mathbb{E}_{\mathbf{x} \sim {p_{\mathbf{x}}}(\mathbf{x})}}[||\mathcal{R}_{\mathbf{x}} - \mathbf{x} |{|_1}] \\
  &{}= {\mathbb{E}_{\mathbf{x} \sim {p_{\mathbf{x}}}(\mathbf{x} )}}[||{G_{\mathbf{y}\rightarrow\mathbf{x} }}({G_{\mathbf{x}\rightarrow\mathbf{y}}}(\mathbf{x}))-\mathbf{x} |{|_1}], \\
  {\mathcal{R}_{\mathbf{rec(y)}}} &{}=  {\mathbb{E}_{\mathbf{y} \sim {p_{\mathbf{y}}}(\mathbf{y} )}}[||\mathcal{R}_{\mathbf{y}}-\mathbf{y} |{|_1}] \\
 &{} ={\mathbb{E}_{\mathbf{y} \sim {p_{\mathbf{y}}}(\mathbf{y} )}}[||{G_{\mathbf{x}\rightarrow\mathbf{y} }}({G_{\mathbf{y}\rightarrow\mathbf{x}}}(\mathbf{y}))-\mathbf{y} |{|_1}],\\
 {\mathcal{R}_{\mathbf{Total}}} &{}= {\mathcal{R}_{\mathbf{rec(x)}}}+ {\mathcal{R}_{\mathbf{rec(y)}}}.
\end{array}
\end{equation}
\subsubsection{Identity-tracing loss}
Linguistic information of speakers plays a vital role in a voice conversion task. Identity-tracing loss is adopted to preserve linguistic information. This loss can enhances similarity measure between the generated speech samples and the target speech samples. So, the identity-tracing losses for the speech samples $\mathbf{x}$ and $\mathbf{y}$ in terms of $L_1$ loss are defined as
\begin{equation}
\begin{array}{l@{}l}
{\mathcal{{I}_{\mathbf{trace(x)}}}} &{}= 
  {\mathbb{E}_{\mathbf{x} \sim {p_{\mathbf{x}}}(\mathbf{x})}}[||\mathcal{I}_{\mathbf{x}} - \mathbf{x} |{|_1}] \\
  &{} = {\mathbb{E}_{\mathbf{x} \sim {p_{\mathbf{x}}}(\mathbf{x})}}[||{G_{\mathbf{y}\rightarrow\mathbf{x} }}(\mathbf{x})-\mathbf{x} |{|_1}], \\
  {\mathcal{{I}_{\mathbf{trace(y)}}}} &{}= 
  {\mathbb{E}_{\mathbf{y} \sim {p_{\mathbf{y}}}(\mathbf{y})}}[||\mathcal{I}_{\mathbf{y}} - \mathbf{y} |{|_1}] \\
   &{}= {\mathbb{E}_{\mathbf{y} \sim {p_{\mathbf{y}}}(\mathbf{Y})}}[||{G_{\mathbf{X}\rightarrow\mathbf{y} }}(\mathbf{y})-\mathbf{y} |{|_1}],\\
 \therefore \quad
  {\mathcal{I}_{\mathbf{Total}}} &{}= {\mathcal{I}_{\mathbf{trace(x)}}}+ {\mathcal{I}_{\mathbf{trace(y)}}}
\end{array} 
\end{equation}
\subsubsection{Full loss-function of our model}
The overall loss function of a model is the summation of the adversarial loss, the reconstruction loss and the identity-tracing loss. Finally, the full loss function of the proposed model denoted as ${\mathcal{L}_{Full}{(ALGAN-VC)}}$ and expressed as
\begin{equation}
  \mathcal{L}_{Full}{(ALGAN-VC)} ={\mathcal{L}_{\mathbf{Total}}}{\text{ }}+ {\mathcal{R}_{\mathbf{Total}}}{\text{}}+{\mathcal{I}_{\mathbf{Total}}}\\ 
\label{eq:14}
\end{equation}

\subsection{Boosted learning rate strategy (BLRS)}
For the efficient and effective training purpose of our model, boosted learning rate strategy is proposed in this study. Mostly, continuous occurrence of large gradient restricts gradient-based parameter optimization process in many DNN model \cite{learning-rate-boosting}. It provides difficulty in choosing the appropriate learning rate in parameter update rule during training of a model. To overcome this issue, the learning rate is updated adaptively (called as boosted) throughout the training of ALGAN-VC model. The pseudo code of the proposed boosted learning rate strategy is presented in Algorithm \ref{algo-3}.
\LinesNumberedHidden
\begin{algorithm}
{Initialization:} Input samples: $\mathbf{X}$ and $\mathbf{Y}$, Learning rate for the generator and discriminator: ${\eta}_{g}$ and ${\eta}_{d}$, Generator loss scale factor: $\lambda$ , constants: $c_1$, $c_2$ and maximum epochs: $N$\\
\While{$i \leq N$}{
  \uIf{$i = 1$}{
   ${G}_{{Loos}_{i}}$, ${D}_{{Loos}_{i}}$ $\gets \mathcal{\textit{ALGAN-VC}}$ ($\mathbf{X}$, $\mathbf{Y}$, ${\eta}_{g}$, ${\eta}_{d}$)\\
   }
   \Else{
  ${G}_{{Loos}_{i}}$, ${D}_{{Loos}_{i}}$ $\gets \mathcal{\textit{ALGAN-VC}}$
   ($\mathbf{X}$, $\mathbf{Y}$, ${\eta}_{g}$, ${\eta}_{d}$)\\
   ${\Delta}_{{G}_{Loss}}$ $=$ \textit{abs}(${G}_{{Loos}_{i}}$ $-$ ${G}_{{Loos}_{i-1}}$) \\
   ${\Delta}_{{D}_{Loss}}$ $=$ \textit{abs}(${D}_{{Loos}_{i}}$ $-$ ${D}_{{Loos}_{i-1}}$) \\
 \uIf{ ${\lambda} \times {\Delta}_{{G}_{Loss}} > {\Delta}_{{D}_{Loss}}$}{
   ${\eta}_{g}$ $=$  ${\eta}_{g}$ $-$ $c_1$, 
   ${\eta}_{d}$ $=$  ${\eta}_{d}$ $+$ $c_2$ 
   }
  \uElseIf{${\lambda} \times {\Delta}_{{G}_{Loss}} < {\Delta}_{{D}_{Loss}}$}{
   ${\eta}_{g}$ $=$  ${\eta}_{g}$ $+$ $c_2$,
   ${\eta}_{d}$ $=$  ${\eta}_{d}$ $-$ $c_1$ 
   }
  \Else{
     ${\eta}_{g}$=${\eta}_{g}$,
     ${\eta}_{d}$=${\eta}_{d}$ 
  }
   }
    $i=i+1$ \\
  }
\caption{Boosted Learning Rate Strategy}
\label{algo-3}
\end{algorithm}

\par
Algorithm \ref{algo-3} start with generator loss (${G}_{Loos}$) and discriminator loss (${D}_{Loos}$) which are obtained from ALGAN-VC model using the initial learning rates (${\eta}_{g}$ and ${\eta}_{d}$) of the generator and discriminator along with initial samples $\mathbf{X}$ and $\mathbf{Y}$ in the first epoch. From the second epoch, the rate of change of losses in both the generator and discriminator (${\Delta}_{{G}_{Loss}}$ and ${\Delta}_{{D}_{Loss}}$) are calculated for updating the learning rates. Here, $\lambda$ scale factor is multiplied with ${\Delta}_{{G}_{Loss}}$ for fair comparison with ${\Delta}_{{D}_{Loss}}$. Finally, the $\eta_g$ and $\eta_d$ are updated with two constants $c_1$ and $c_2$ depending on the loss increments or decrements of the generator and the discriminator in consecutive epochs.

\section{Experimental Setups} \label{exp-setup}
\subsection{Dataset Description}
For evaluating the performance of the proposed ALGAN-VC model, the Voice Conversion Challenge (VCC) $2016$ \cite{2016}, VCC $2018$ \cite{2018}, VCC $2020$ speech dataset \cite{2020} and our self prepared nonparallel Indian multilingual datasets{\footnote{Self prepared dataset is available at \url{https://tinyurl.com/sjyjzebk}}} are used in the experiment. The VCC $2016$ parallel speech dataset was recorded by $8$ speakers in US English and segmented into $216$ utterances. The whole dataset is divided into source and target speakers' data contained $2$ male and $2$ female speakers in each division. The VCC $2018$ speech data is recorded by $12$ US English speakers and divided into parallel and non-parallel datasets with total $116$ utterances. Both the parallel and non-parallel divisions of the VCC $2018$ dataset are grouped into source and target speaker's data with $2$ male and $2$ female speakers. Unlike the VCC $2016$ and the VCC $2018$ monolingual dataset, the VCC $2020$ consists of both monolingual and crosslingual speech data. The monolingual speech dataset of the VCC $2020$ is divided into $20$ parallel and $50$ nonparallel utterances, collected from $4$ source and $4$ target speakers in English.  For the crosslingual speech dataset of the VCC $2020$,  English, Finnish, German and Mandarin languages are considered with $70$ nonparallel utterances. Moreover, we prepared a set of non-parallel datasets of $15$ speakers, each containing $20$ instances with $1715$ utterances. Among them, $10$ speakers' speeches are recorded in the English language, and the remaining $5$ speakers' speeches in $5$ different Indian regional languages such as Bengali, Hindi, Assamese, Odia, and Malayalam in both male and female voice. The summary of the experimented datasets are presented in Table \ref{table-dataset}.
\begin{table*}
\caption{SUMMARY OF THE DATASETS} 
\centering 
\renewcommand{\arraystretch}{1.5}
\resizebox{0.98\textwidth}{!}{%
\begin{tabular}{ccccc}  
\hline\hline 
Dataset & Type & Language & Utterances & Time interval (in seconds) \\[1ex]
\hline 
VCC 2016 & Parallel & US English & 216 & 3-7 \\ \hline
\multirow{2}{*}{VCC 2018}  & Parallel     & US English      & 116 & 3-9 \\ 
                           & Non Parallel & US English       & 116 & 3-9 \\ \hline
\multirow{2}{*}{VCC  2020} & Parallel     & US English      & 20  & 4-7 \\ 
                           & Non Parallel & US English, German, Finnish, Mandarin & 120 & 3-7 \\ \hline
Self-prepared & Non Parallel &  Indian accent of English and Indian regional languages (Bengali, Assamese, Hindi, Odia, Malayalam) & 1715 & 5-9 \\ \hline\hline                       
\end{tabular}
}
\label{table-dataset} 
\end{table*}

\subsection{Feature Extraction}
For the experimental purpose, each dataset is downsampled to $16$KHz with a bit depth of $16$ bits, respectively. The three widely used important speech features such as Mel-cepstral coefficient (MCEP), Aperiodicities (AP), and Logarithmic fundamental frequency (log${F}_{0}$) are considered in our study for voice conversion task. These features are extracted with the help of the WORLD \cite{world} vocoder feature analysis system. The MCEP features are used in the generator network for speech feature mapping from source speaker to target speaker while ${F}_{0}$ conversion is performed using logarithm Gaussian normalized transformation \cite{Gaussian}. Additionally, AP features are converted directly from source speaker to target speaker without any modification \cite{APS}. Finally, speech features are reconstructed using the WORLD vocoder system.

\subsection{Network Architecture of ALGAN-VC}
In the proposed ALGAN-VC model, the network architecture framework is adopted from a GAN model used in image style transfer \cite{CycleGAN2017}. The overall network architecture of ALGAN-VC can be visualized in Fig. \ref{basic-frame-work}. The generator network initially connected with two consecutive convolution blocks, each of block consists 1D CNN and GLU activation function. Here, 1D CNN is used because it can capture the overall relationship among the input features and the gated mechanism of GLU allows selective propagation of information that depends on the previous layers outputs. Furthermore, inspired from the effective use of downsampling blocks, residual blocks and upsampling blocks as well as instance normalization (IN) in image style transfer research \cite{STyle-transfer, IN}, similar block sequence is employed in ALGAN-VC generator for adequate speech features transfer. In the sequential downsampling and upsampling blocks, sequential upscaling and downscaling of channel numbers are used for capturing fine distribution of data more explicitly. In our study, eight dense residual blocks are used with five downsampling and upsampling blocks in the generator network. Additionally, due to the effectiveness of pixel shuffler for high-resolution image generation task \cite{cvpr}, it is incorporated in upsampling blocks. Finally, the generator network produced output as converted speech features. The generator network architecture settings are presented in Table \ref{generator-table}.
\begin{table}
\caption{ALGAN-VC GENERATOR NETWORK ARCHITECTURE} 
\centering 
\renewcommand{\arraystretch}{1.5}
\resizebox{0.4\textwidth}{!}{%
\begin{tabular}{ccccc}  
\hline\hline 
Conv Blocks & Channel & Kernal & Stride & Output \\
[0.5ex]
\hline 
Input Conv &   24 &   1x5 &   1x1 &   24 \\
Input Conv &   64 &  1x5 &  1x1 &  64 \\ \hline
 Downsampling &  64 &  1x5 &  1x2 &  1024 \\ 
 Downsampling &  128 &  1x5 &  1x2 &  1024 \\
 Downsampling &  256 &  1x5 &  1x2 &  1024 \\
 Downsampling &  512 &  1x5 &  1x2 &  1024 \\
 Downsampling &  1024 &  1x5 &  1x2 &  1024 \\ \hline
 Dense Residual blocks &   1024 &  1x5 &  1x1 &  1024  \\  \hline
 Upsampling &  64 &  1x5 &  1x1 &  1024 \\ 
 Upsampling &  128 &  1x5 &  1x1 &  1024 \\
 Upsampling &  256 &  1x5 &  1x1 &  1024 \\
 Upsampling &  512 &  1x5 &  1x1 &  1024 \\
 Upsampling &  1024 &  1x5 &  1x1 &  1024 \\ \hline
 Output Conv &  24 &  1x5 &  1x1 &  24 \\
\hline\hline 
\end{tabular}
}
\label{generator-table} 
\end{table}
\par
The discriminator network of the ALGAN-VC is a 2D CNN model comprise of two input convolution blocks and five downsampling blocks. The advantage of 2D CNN is, it can capture the details of time-frequency representation (i.e spectral texture) and thus widely used for spectral classification \cite{2DCNN}. The details of the discriminator network architecture such as channel number, kernel size, stride and output are given in Table \ref{discriminator-table}. 
\begin{table}
\caption{ALGAN-VC DISCRIMINATOR NETWORK ARCHITECTURE} 
\centering
\renewcommand{\arraystretch}{1.5}
\resizebox{0.4\textwidth}{!}{%
\begin{tabular}{ccccc}  
\hline\hline 
Conv Blocks & Channel & Kernal & Stride & Output \\
[0.5ex]
\hline 
Input Conv &  24 &  4x4 &  1x2 &  24 \\
Input Conv &  64 &  4x4 &  1x2 &  64 \\ \hline
Downsampling &  64 &  4x4 &  1x2 &  1024 \\ 
Downsampling &  128 &  4x4 &  1x2 &  1024 \\
Downsampling &  256 &  4x4 &  1x2 &  1024 \\
Downsampling &  512 &  4x4 &  1x2 &  1024 \\
Downsampling &  1024 &  4x4 &  1x2 &  1024 \\
\hline\hline 
\end{tabular}
}
\label{discriminator-table} 
\end{table}
\subsection{Training Details}
All the experiments are conducted by considering $24$ MCEPs. The initial learning rate of the generator and the discriminator are set as $0.0002$ and $0.0001$, respectively. In Algorithm \ref{algo-3}, the constants $c_1$ and $c_2$ are set as ${1\times10^{-6}}$ and ${1\times10^{-5}}$. It is observed in boosted learning rate strategy that ${{\Delta}_{{G}_{Loss}}>>{\Delta}_{{D}_{Loss}}}$ in consecutive epochs. Thus, a scale factor ${\lambda}$ is used to scale ${{\Delta}_{{G}_{Loss}}}$ and set the value as ${5\times10^{-2}}$. The Adam optimizer is used to trained our proposed ALGAN-VC model. The momentum term for this optimizer is $0.5$. Throughout the experiment, the batch size is considered as $1$ while mini-batches are constructed from $128$ randomly selected frames. In our study, the termination condition is the maximum number of epochs which is set as ${1.5\times10^{4}}$ epochs.

\subsection{Experimental Setup}
The ALGAN-VC model and all the experiments are implemented in Python $3.6.9$ using two libraries as Tensorflow $1.15.0$ and Keras $2.3.1$. Audio data is prepossessed by using Librosa $0.7.2$ and Pyworld $0.2.8$. All the speech features are stored in the form of {\textit{.npz}} format using the library Numpy $1.15$. Finally, the whole processes is executed in Dell precision $7820$ workstation configured with ubuntu $18.04$ $64$ bit Operating System, Intel Xeon Gold 5215 2.5GHz processor, 96GB RAM and Nvidia 16GB Quadro RTX5000 graphics.

\section{Results and Discussion} \label{result}
The performance of the proposed ALGAN-VC model is assessed with subjective evaluations and objective evaluations. Our results are compared with CycleGAN-VC \cite{cyclegan-vc-1}, CycleGAN-VC2 \cite{cyclegan-vc-2}, and Spectrum-Prosody-CycleGAN (SP-CycleGAN) \cite{cyclegan-vc-7} with respect to (w.r.t) two evaluations to measured the effectiveness of the proposed model. Additionally, ablation study is performed for the ALGAN-VC model w.r.t DRN blocks, learning rate and loss functions.  
\subsection{Subjective Evaluations}
The ABX test \cite{abx-test} and Mean Opinion Score (MOS) \cite{mos} are used for subjective evaluations to measured the speaker similarity and speech quality for the converted speech samples. These scores are collected by employing $20$ volunteers for both intra gender and inter gender converted speech samples of our experimental datasets. The ABX test and MOS results of the proposed model and the compared models as well as the ablation study over the datasets are presented in Table \ref{ABX-TEST-Table} and Table \ref{MOS}.
\begin{table*}
\centering
\renewcommand{\arraystretch}{1.2}
\caption{RESULTS OF ABX TEST BETWEEN THE PROPOSED ALGAN-VC MODEL AND OTHER COMPARED MODELS FOR SPEAKER SIMILARITY ( IN \%) }
\resizebox{1.01\textwidth}{!}{%
\begin{tabular}{ccccccccccc}
\hline\hline
Dataset &
  \multicolumn{2}{c}{Type of data} &
  \begin{tabular}[c]{@{}c@{}}CycleGAN-VC\end{tabular} &
  \begin{tabular}[c]{@{}c@{}}CycleGAN-VC2\end{tabular} &
  \begin{tabular}[c]{@{}c@{}}SP-CycleGAN\end{tabular} &
  ALGAN-VC &
 \begin{tabular}[c]{@{}c@{}} ALGAN-VC \\ without DRN \end{tabular} &
 \begin{tabular}[c]{@{}c@{}} ALGAN-VC \\ without BLRS \end{tabular}&
  \begin{tabular}[c]{@{}c@{}}ALGAN-VC\\ without L1 loss\end{tabular} &
  \begin{tabular}[c]{@{}c@{}}ALGAN-VC\\ without L2 loss\end{tabular} \\ \hline\hline
\multirow{5}{*}{\begin{tabular}[c]{@{}c@{}}VCC 2016\\ parallel data\end{tabular}} &
  \multirow{2}{*}{\begin{tabular}[c]{@{}c@{}}Intra gender\\ confidence score\end{tabular}} &
  M-M &  67.50 &  72.50 &  70.00 &  77.50 &  65.50 &  65.50 &  70.00 &  65.50 \\
 &
   &
  F-F &
  65.00 &
  70.00 &
  70.00 &
  72.50 &
  67.00 &
  62.50 &
  67.50 &
  67.50 \\ \cline{2-11} 
 &
  \multirow{2}{*}{\begin{tabular}[c]{@{}c@{}}Inter gender\\ confidence score\end{tabular}} &
  M-F &
  55.50 &
  61.87 &
  57.00 &
  62.50 &
  52.50 &
  55.50 &
  62.50 &
  52.50 \\
 &
   &
  F-M &
  55.50 &
  55.50 &
  55.00 &
  65.00 &
  55.00 &
  50.00 &
  55.50 &
  52.50 \\ \cline{2-11} 
 &
  \multicolumn{2}{c}{Average confidence score} &
  {60.87} &
  {64.96} &
  {63.12} &
 \textbf{69.37} &
  {60.00} &
  {58.25} &
  {63.87} &
  {60.75} \\ \hline
\multirow{5}{*}{\begin{tabular}[c]{@{}c@{}}VCC 2018 \\ non parallel data\end{tabular}} &
  \multirow{2}{*}{\begin{tabular}[c]{@{}c@{}}Intra gender \\ confidence score\end{tabular}} &
  M-M &
  60.00 &
  67.50 &
  65.00 &
  70.00 &
  67.50 &
  60.00 &
  67.50 &
  60.00 \\
 &
   &
  F-F &
  65.00 &
  65.00 &
  67.50 &
  67.50 &
  65.00 &
  57.50 &
  67.50 &
  65.00 \\ \cline{2-11} 
 &
  \multirow{2}{*}{\begin{tabular}[c]{@{}c@{}}Inter gender\\ confidence score\end{tabular}} &
  M-F &
  47.50 &
  55.50 &
  52.50 &
  57.50 &
  50.00 &
  47.50 &
  55.50 &
  50.00 \\
 &
   &
  F-M &
  50.00 &
  50.50 &
  52.50 &
  60.00 &
  55.50 &
  50.00 &
  55.50 &
  55.50 \\ \cline{2-11} 
 &
  \multicolumn{2}{c}{Average confidence score} &
  55.62 &
  59.50 &
  59.37&
  \textbf{63.75} &
  59.50 &
  53.75 &
  61.50 &
  57.62 \\ \hline
\multirow{5}{*}{\begin{tabular}[c]{@{}c@{}}VCC 2020\\mono lingual non parallel data\end{tabular}} &
  \multirow{2}{*}{\begin{tabular}[c]{@{}c@{}}Intra gender\\ confidence score\end{tabular}} &
  M-M &
  60.00 &
  65.00 &
  67.50 &
  67.50 &
  62.50 &
  60.00 &
  67.50 &
  57.50 \\
 &
   &
  F-F &
  60.00 &
  60.00 &
  65.00 &
  67.50 &
  65.00 &
  62.50 &
  67.50 &
  62.50 \\ \cline{2-11} 
 &
  \multirow{2}{*}{\begin{tabular}[c]{@{}c@{}}Inter gender\\ confidence score\end{tabular}} &
  M-F &
  50.00 &
  50.00 &
  55.00 &
  55.00 &
  47.50 &
  55.00 &
  55.00 &
  47.50 \\
 &
   &
  F-M &
  50.00 &
  55.00 &
  55.00 &
  57.50 &
  50.00 &
  45.00 &
  50.00 &
  47.50 \\ \cline{2-11} 
 &
  \multicolumn{2}{c}{Average confidence score} &
  {55.00} &
  {57.62} &
  {60.62} &
  \textbf{61.87} &
  {56.25} &
  {55.62} &
  {60.00} &
  {53.75} \\ \hline
\multirow{5}{*}{\begin{tabular}[c]{@{}c@{}}VCC 2020\\cross lingual non parallel data\end{tabular}} &
  \multirow{2}{*}{\begin{tabular}[c]{@{}c@{}}Intra gender\\ confidence score\end{tabular}} &
  M-M &
  55.00 &
  55.00 &
  55.00 &
  62.50 &
  57.50 &
  55.00 &
  60.00 &
  55.00 \\
 &
   &
  F-F &
  50.00 &
  60.00 &
  55.00 &
  60.00 &
  60.00 &
  60.50 &
  62.50 &
  60.00 \\ \cline{2-11} 
 &
  \multirow{2}{*}{\begin{tabular}[c]{@{}c@{}}Inter gender\\ confidence score\end{tabular}} &
  M-F &
  45.00 &
  50.00 &
  50.00 &
  50.00 &
  47.50 &
  47.50 &
  50.00 &
  55.00 \\
 &
   &
  F-M &
  47.50 &
  47.50 &
  50.00 &
  50.00 &
  55.00 &
  47.00 &
  50.00 &
  45.00 \\ \cline{2-11} 
 &
  \multicolumn{2}{c}{Average confidence score} &
  {49.37} &
  {53.12} &
  {52.50} &
  \textbf{55.62} &
  {55.00} &
  {52.50} &
  \textbf{55.62} &
  {53.75} \\ \hline

\multirow{5}{*}{\begin{tabular}[c]{@{}c@{}}Self prepared non parallel\\ english language data\end{tabular}} &
  \multirow{2}{*}{\begin{tabular}[c]{@{}c@{}}Intra gender\\ confidence score\end{tabular}} &
  M-M &
  55.00 &
  60.00 &
  67.50 &
  67.50 &
  62.50 &
  60.00 &
  67.50 &
  60.00 \\
 &
   &
  F-F &
  60.00 &
  65.00 &
  65.00 &
  67.50 &
  67.50 &
  62.50 &
  62.50 &
  62.50 \\ \cline{2-11} 
 &
  \multirow{2}{*}{\begin{tabular}[c]{@{}c@{}}Inter gender\\ confidence score\end{tabular}} &
  M-F &
  50.00 &
  50.00 &
  55.00 &
  55.50 &
  50.00 &
  50.00 &
  55.00 &
  50.00 \\
 &
   &
  F-M &
  50.00 &
  50.00 &
  50.00 &
  55.00 &
  55.00 &
  50.00 &
  55.00 &
  50.00 \\ \cline{2-11} 
 &
  \multicolumn{2}{c}{Average confidence score} &
  {53.75} &
  {56.25} &
  {59.35} &
  \textbf{61.25} &
  {58.75} &
  {55.50} &
  {60.00} &
  {55.50} \\ \hline
\multirow{5}{*}{\begin{tabular}[c]{@{}c@{}}Self prepared non parallel\\ regional language data\end{tabular}} &
  \multirow{2}{*}{\begin{tabular}[c]{@{}c@{}}Intra gender\\ confidence score\end{tabular}} &
  M-M &
  50.00 &
  60.00 &
  65.00 &
  65.00 &
  62.50 &
  60.00 &
  60.00 &
  60.00 \\
 &
   &
  F-F &
  55.00 &
  55.00 &
  65.00 &
  65.00 &
  65.00 &
  50.00 &
  65.00 &
  60.00 \\ \cline{2-11} 
 &
  \multirow{2}{*}{\begin{tabular}[c]{@{}c@{}}Inter gender\\ confidence score\end{tabular}} &
  M-F &
  47.50 &
  50.00 &
  50.00 &
  50.00 &
  50.00 &
  50.00 &
  55.00 &
  50.00 \\
 &
   &
  F-M &
  50.00 &
  50.00 &
  50.00 &
  55.00 &
  50.50 &
  50.50 &
  50.50 &
  50.00 \\ \cline{2-11} 
 &
  \multicolumn{2}{c}{Average confidence score} &
  {50.62} &
  {53.75} &
  {57.50} &
  \textbf{58.75} &
  {56.87} &
  {52.50} &
  {57.50} &
  {55.00} \\ \hline\hline
\end{tabular}%
}
\label{ABX-TEST-Table}
\end{table*}

\begin{table*}
\centering
\renewcommand{\arraystretch}{1.2}
\caption{RESULTS OF MOS BETWEEN THE PROPOSED ALGAN-VC MODEL AND OTHER COMPARED MODELS FOR SPEECH QUALITY }
\resizebox{1.01\textwidth}{!}{%
\begin{tabular}{ccccccccccc}
\hline\hline
Dataset &
  \multicolumn{2}{c}{Type of data} &
  \begin{tabular}[c]{@{}c@{}}CycleGAN-VC\end{tabular} &
  \begin{tabular}[c]{@{}c@{}}CycleGAN-VC2\end{tabular} &
  \begin{tabular}[c]{@{}c@{}}SP-CycleGAN\end{tabular} &
  ALGAN-VC &
 \begin{tabular}[c]{@{}c@{}} ALGAN-VC \\ without DRN \end{tabular} &
 \begin{tabular}[c]{@{}c@{}} ALGAN-VC \\ without BLRS \end{tabular}&
  \begin{tabular}[c]{@{}c@{}}ALGAN-VC\\ without L1 loss\end{tabular} &
  \begin{tabular}[c]{@{}c@{}}ALGAN-VC\\ without L2 loss\end{tabular} \\ \hline\hline
\multirow{5}{*}{\begin{tabular}[c]{@{}c@{}}VCC 2016\\ parallel data\end{tabular}} &
  \multirow{2}{*}{\begin{tabular}[c]{@{}c@{}}Intra gender\\ confidence score\end{tabular}} &
  M-M &
  1.98 &
  1.55 &
  2.03 &
  2.10 &
  1.87 &
  1.78 &
  1.57 &
  1.69 \\
 &
   &
  F-F &
  1.78 &
  1.98 &
  1.98 &
  2.01 &
  1.98 &
  1.88 &
  1.69 &
  1.55\\ \cline{2-11} 
 &
  \multirow{2}{*}{\begin{tabular}[c]{@{}c@{}}Inter gender\\ confidence score\end{tabular}} &
  M-F &
  1.03 &
  1.31 &
  1.57 &
  1.67 &
  1.13 &
  1.32 &
  1.38 &
  1.32 \\
 &
   &
  F-M &
  1.13 &
  1.57 &
  1.78 &
  1.98 &
  1.13 &
  1.78 &
  1.78 &
  1.11 \\ \cline{2-11} 
 &
  \multicolumn{2}{c}{Average confidence score} &
  {1.48} &
  {1.60} &
  {1.84} &
 \textbf{1.94} &
  {1.52} &
  {1.69} &
  {1.60} &
  {1.41} \\ \hline
\multirow{5}{*}{\begin{tabular}[c]{@{}c@{}}VCC 2018 \\ non parallel data\end{tabular}} &
  \multirow{2}{*}{\begin{tabular}[c]{@{}c@{}}Intra gender \\ confidence score\end{tabular}} &
  M-M &
  1.53 &
  1.68 &
  1.87 &
  1.78 &
  1.66 &
  1.57 &
  1.51 &
  1.61 \\
 &
   &
  F-F &
  1.39 &
  1.73 &
  1.91 &
  1.89 &
  1.73 &
  1.69 &
  1.49 &
  1.57 \\ \cline{2-11} 
 &
  \multirow{2}{*}{\begin{tabular}[c]{@{}c@{}}Inter gender\\ confidence score\end{tabular}} &
  M-F &
  1.17 &
  1.15 &
  1.23 &
  1.27 &
  1.09 &
  1.13 &
  1.32 &
  1.13 \\
 &
   &
  F-M &
  1.15 &
  1.23 &
  1.57 &
  1.63 &
  1.17 &
  1.19 &
  1.55 &
  1.12 \\ \cline{2-11} 
 &
  \multicolumn{2}{c}{Average confidence score} &
  1.31 &
  1.44 &
  \textbf{1.64}&
  \textbf{1.64} &
  1.41&
  1.39 &
  1.46 &
  1.35 \\ \hline
\multirow{5}{*}{\begin{tabular}[c]{@{}c@{}}VCC 2020\\mono lingual non parallel data\end{tabular}} &
  \multirow{2}{*}{\begin{tabular}[c]{@{}c@{}}Intra gender\\ confidence score\end{tabular}} &
  M-M &
  1.23 &
  1.77 &
  1.91 &
  1.88 &
  1.77 &
  1.32 &
  1.75 &
  1.62 \\
 &
   &
  F-F &
  1.57 &
  1.83 &
  1.67 &
  1.98 &
  1.73 &
  1.27 &
  1.73 &
  1.51 \\ \cline{2-11} 
 &
  \multirow{2}{*}{\begin{tabular}[c]{@{}c@{}}Inter gender\\ confidence score\end{tabular}} &
  M-F &
  1.12 &
  1.32 &
  1.39 &
  1.32 &
  1.51 &
  1.21 &
  1.45 &
  1.32 \\
 &
   &
  F-M &
  1.17 &
  1.37 &
  1.42 &
  1.37 &
  1.53 &
  1.17 &
  1.39 &
  1.39 \\ \cline{2-11} 
 &
  \multicolumn{2}{c}{Average confidence score} &
  {1.27} &
  {1.57} &
  {1.59} &
  \textbf{1.63} &
  \textbf{1.63} &
  {1.24} &
  {1.58} &
  {1.46} \\ \hline
\multirow{5}{*}{\begin{tabular}[c]{@{}c@{}}VCC 2020\\cross lingual non parallel data\end{tabular}} &
  \multirow{2}{*}{\begin{tabular}[c]{@{}c@{}}Intra gender\\ confidence score\end{tabular}} &
  M-M &
  1.27 &
  1.68 &
  1.78 &
  1.87 &
  1.73 &
  1.39 &
  1.69 &
  1.57 \\
 &
   &
  F-F &
  1.39 &
  1.57 &
  1.69 &
  1.78 &
  1.67 &
  1.23 &
  1.55 &
  1.32 \\ \cline{2-11} 
 &
  \multirow{2}{*}{\begin{tabular}[c]{@{}c@{}}Inter gender\\ confidence score\end{tabular}} &
  M-F &
  1.23 &
  1.15 &
  1.31 &
  1.39 &
  1.23 &
  1.45 &
  1.11 &
  1.17 \\
 &
   &
  F-M &
  1.12 &
  1.23 &
  1.27 &
  1.31 &
  1.57 &
  1.51 &
  1.32 &
  1.15 \\ \cline{2-11} 
 &
  \multicolumn{2}{c}{Average confidence score} &
  {1.25} &
  {1.40} &
  {1.51} &
  \textbf{1.58} &
  {1.55} &
  {1.39} &
  {1.41} &
  {1.30} \\ \hline

\multirow{5}{*}{\begin{tabular}[c]{@{}c@{}}Self prepared non parallel\\ english language data\end{tabular}} &
  \multirow{2}{*}{\begin{tabular}[c]{@{}c@{}}Intra gender\\ confidence score\end{tabular}} &
  M-M &
  1.57 &
  1.66 &
  1.73 &
  1.87 &
  1.61 &
  1.51 &
  1.49 &
  1.32 \\
 &
   &
  F-F &
  1.45 &
  1.73 &
  1.78 &
  1.73 &
  1.69 &
  1.62 &
  1.69 &
  1.55 \\ \cline{2-11} 
 &
  \multirow{2}{*}{\begin{tabular}[c]{@{}c@{}}Inter gender\\ confidence score\end{tabular}} &
  M-F &
  1.23 &
  1.23 &
  1.32 &
  1.23 &
  1.12 &
  1.17 &
  1.55 &
  1.27 \\
 &
   &
  F-M &
  1.15 &
  1.57 &
  1.55 &
  1.39 &
  1.13 &
  1.21 &
  1.32 &
  1.23 \\ \cline{2-11} 
 &
  \multicolumn{2}{c}{Average confidence score} &
  {1.38} &
  {1.54} &
  \textbf{1.59} &
  {1.55} &
  {1.38} &
  {1.37} &
  {1.51} &
  {1.34} \\ \hline
\multirow{5}{*}{\begin{tabular}[c]{@{}c@{}}Self prepared non parallel\\ regional language data\end{tabular}} &
  \multirow{2}{*}{\begin{tabular}[c]{@{}c@{}}Intra gender\\ confidence score\end{tabular}} &
  M-M &
  1.53 &
  1.75 &
  1.78 &
  1.91 &
  1.69 &
  1.32 &
  1.75 &
  1.55 \\
 &
   &
  F-F &
  1.57 &
  1.73 &
  1.87 &
  1.73 &
  1.62 &
  1.49 &
  1.69 &
  1.66 \\ \cline{2-11} 
 &
  \multirow{2}{*}{\begin{tabular}[c]{@{}c@{}}Inter gender\\ confidence score\end{tabular}} &
  M-F &
  1.27 &
  1.39 &
  1.63 &
  1.51 &
  1.21 &
  1.12 &
  1.32 &
  1.17 \\
 &
   &
  F-M &
  1.15 &
  1.45 &
  1.27 &
  1.55 &
  1.15 &
  1.17 &
  1.37 &
  1.13 \\ \cline{2-11} 
 &
  \multicolumn{2}{c}{Average confidence score} &
  {1.38} &
  {1.58} &
  {1.63} &
  \textbf{1.67} &
  {1.41} &
  {1.27} &
  {1.53} &
  {1.37} \\ \hline\hline
\end{tabular}%
}
\label{MOS}
\end{table*}
\par
ABX test score or confidence score is collected based on eight randomly selected converted speech samples provided to each of the volunteers. For evaluating the speaker similarity, score is assigned by the values $1$ and $0$ for correct and incorrect identification of speakers. Table \ref{ABX-TEST-Table} shown the average of the collected scores for each of the models in percentage. The confidence score of intra gender is denoted as M-M (male to male) or F-F (female to female) while inter gender is denoted as M-F (male to female) or F-M (female to male) for their converted speech data. For VCC 2016 parallel speech data, ALGAN-VC scored highest with $69.37$\% speaker similarity score which reflects the effectiveness of the proposed model for mono lingual parallel voice conversion. Moreover for VCC 2018, VCC 2020 mono lingual non parallel speech data, exhibited the superiority of our proposed model with $63.75\%$ and $61.87\%$ speaker similarity score compared to other models. Although, CycleGAN-VC2 and SP-CycleGAN placed second position with $59.50\%$ and $60.62$\% speaker similarity score for VCC 2018 and VCC 2020 mono lingual non parallel speech conversion (without considering the ablation study results), respectively. Moreover, it can be observed that ALGAN-VC performed well for cross lingual voice conversion in case of VCC 2020 cross lingual speech data compared to CycleGAN-VC, CycleGAN-VC2 and SP-CycleGAN. ALGAN-VC also showed better results compared to other models over the self-prepared speech data besides the standard speech data. From Table \ref{ABX-TEST-Table}, it can be noticed that the speaker similarity score is more for parallel voice conversion than the non parallel voice conversion as the feature mapping is simpler in case of parallel speech data. Moreover, the confidence scores for M-M and F-F are relatively higher than M-F and F-M. It is because, the Logarithmic fundamental frequency (${log{F}_{0}}$) of male and female speakers are vastly different which results in adequate conversion of inter gender ${log{F}_{0}}$ features. Therefore, the proposed model provides the significant improvements in intra gender voice conversion than inter gender. It is also well observed that low speaker similarity scores are obtained in cross lingual voice conversion data compared to mono lingual voice voice conversion data irrespective of models. Hence, the performance is degraded due to the presence of unaligned corpus in cross lingual speech data. However, our proposed model reveals the better performance compared to other models over the cross lingual voice conversion data because of well feature learning mapping ability. Although, effective feature mapping is necessary to achieve high speaker similarity score in case of cross lingual voice conversion.
\par
Also, ablation study is performed on the proposed ALGAN-VC model and their corresponding results of ABX test scores over the datasets are shown in Table \ref{ABX-TEST-Table}. Firstly removed the DRN like generator architecture from the ALGAN-VC model and used only residual connection based generator architecture called as ALGAN-VC without DRN. This degrades the performance of the model marginally in terms of speaker similarity. Secondly, BLRS is replaced with constant learning rate approach in ALGAN-VC and observed that it results significant degradation in speaker similarity scores form ALGAN-VC model. Thirdly, we investigated ALGAN-VC models' performance by removing $L_1$ loss from the linear combinations of loss functions called as ALGAN-VC without $L_1$ loss. It can be observed that this strategy provides speaker similarity scores which are very closer to the original ALGAN-VC model in case of VCC 2020 and self prepared datasets. This signifies the utility of $L_2$ loss in the linear combination of loss functions. Finally, ALGAN-VC models' performance is investigated without $L_2$ loss and degraded the speaker similarity score results as compared to the main ALGAN-VC model. So, $L_2$ loss is very important for the higher score.
\par
On the other hand, MOS is collected from the volunteers to evaluate the models' performance in terms of speech quality. In a similar manner like ABX test, randomly selected speech samples are presented to each of the volunteers for scoring the speech samples. In MOS score, each volunteer is asked to score the speech samples with in a range between $1$ to $5$. The average of the collected MOS for the proposed model and the compared models over the various complex speech datasets are shown in Table \ref{MOS}. From Table \ref{MOS}, it can be observed that proposed model scored highest MOS score with the value $1.94$ for the VCC 2016 parallel speech data. In case of VCC 2018 non parallel data, both ALGAN-VC and SP-CycleGAN are attain the same MOS score value $1.64$. Though, ALGAN-VC model exhibited competitive performance for VCC 2020 mono lingual and cross lingual non parallel data and self prepared non parallel regional language data as compared to other models. Only SP-CycleGAN achieved higher MOS score from ALGAN-VC on our self prepared non parallel English speech data with negligible difference $0.04$. Finally, the over all performance of the proposed model is superior than the other models in terms of speech quality for the experimental datasets.
\par
In the ablation study of ALGAN-VC model interms of speech quality it is noticeable that ALGAN-VC without DRN performed significantly well. Moreover for VCC 2020 mono lingual non parallel speech data it obtained highest MOS. Although from Table \ref{MOS} it can be well observed that ALGAN-VC without DRN also showed significant performance. Although the overall speech quality scores for all the models as well as ALGAN-VC is considerably less. This turns out to be as models' limitation.

\subsection{Objective Evaluation}
\begin{figure}
\centering
\subfloat[Speaker similarity measure of ALGAN-VC generated speech samples w.r.t original speech samples]{\includegraphics[width = 2.65in]{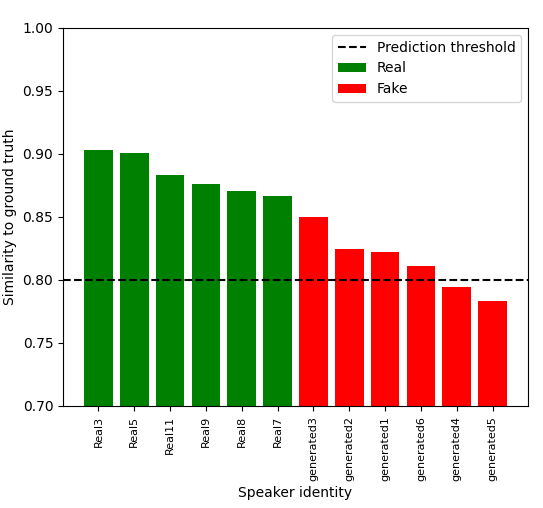}}\\
\subfloat[Speaker similarity measure of SP-CycleGAN generated speech samples w.r.t original speech samples]{\includegraphics[width = 2.65in]{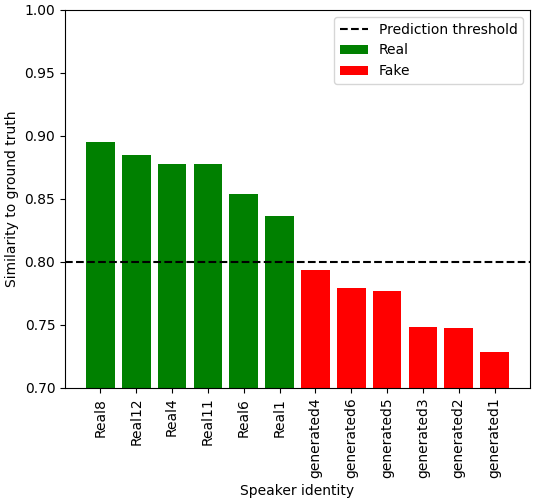}}
\caption{Objective evaluation between ALGAN-VC generated speech data and SP-CycleGAN generated speech data w.r.t original speech data in terms of speaker similarity}
\label{Rysemblyzer}
\end{figure}

The objective evaluation of the generated speech samples responsible about the modest detection of fake speech by comparing the similarity of utterances against the ground truth reference audio data. This evaluation is done by using Rresemblyzer \cite{resemble-ai}. In resemblyzer, a pre-trained vocoder is used to classify real and converted or generated speech samples w.r.t the original speech samples of corresponding speakers treated as ground truth. Based on the outcomes of subjective evaluations, SP-CycleGAN performed second best model compared to other models in terms of speaker similarity and speech quality. So, we performed the objective evaluations of our proposed model in comparison with SP-CycleGAN. The objective evaluations of ALGAN-VC and SP-CycleGAN is shown in Fig. \ref{Rysemblyzer}.
For our experiment, six randomly selected real and generated speech samples are considered and the threshold is set to a standard vale $0.80$. The real and generated speech samples are represented with green and red colour in Fig. \ref{Rysemblyzer}, respectively. It can be observed that the converted speech samples of ALGAN-VC are identified as more real (read colour speeches are above or very closer to $0.80$) than SP-CyleGAN model generated speech samples. This reflects that the generated speech samples using our model is more similar in terms of speaker similarity than SP-CyleGAN model. 
\par
\begin{figure}
\centering
\subfloat[Speaker embeddings for \\ ALGAN-VC generated speech \\ samples]{%
  \includegraphics[height=4.5cm,width=4.7cm]{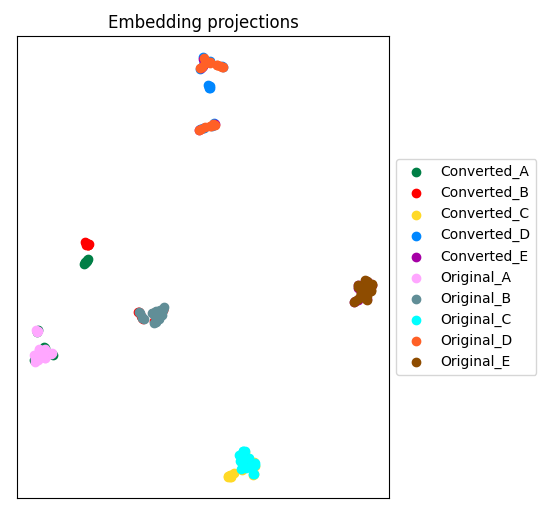}%
  \label{fig:evaluation:revenue1}%
}
\subfloat[Speaker embeddings for \\ SP-CycleGAN generated speech \\ samples]{%
  \includegraphics[height=4.45cm,width=4.3cm]{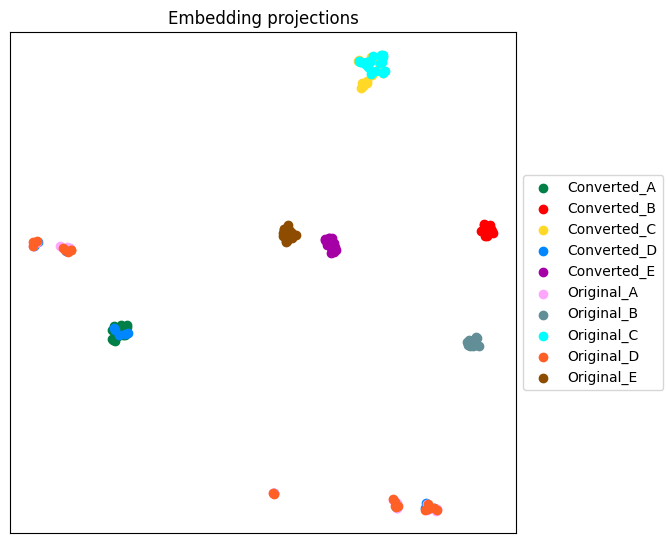}%
  \label{fig:evaluation:avgPrice}%
}
\caption{Projection of speaker embeddings for objective evaluation of ALGAN-VC and SP-CycleGAN model}
\label{embeddings}
\end{figure}

Moreover, the objective evaluation is performed by projecting speaker embeddings. The projection of the speaker embeddings for the two models ALGAN-VC and SP-CycleGAN is presented in Fig. \ref{embeddings},
where each point represents an utterance and the legend indicates different speakers. For ALGAN-VC model, the utterances from the same speakers for both real and generated speech samples are very similar and form a tight cluster while the overlapping clusters signify high similarity between real speech samples and the generated speech samples for the same speakers. The widely scattered clusters of utterances imply the speaker embeddings of different speakers. In case of SP-CycleGAN model, the wider clusters of utterances for the same speakers indicate the vast dissimilarity between the two categories of samples. Also, it can be observed that ALGAN-VC speaker embeddings for original and converted speech samples of respective speakers (denoted as A, B, C, D, E) appeared closer as compare to SP-CycleGAN model. This visualization confirmed that the proposed model can generate better speaker-invariant linguistic representations.
\par
\begin{figure*}
\centering
\subfloat[Mel-spectrogram of original \\ speech considered for SP-CycleGAN]{%
  \includegraphics[width=0.255\textwidth]{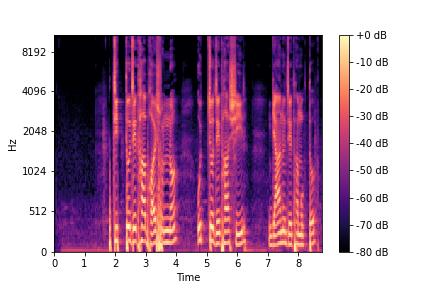}%
  \label{fig:evaluation:revenue}%
  }
\subfloat[Mel-spectrogram of converted \\ speech using SP-CycleGAN]{%
  \includegraphics[width=0.255\textwidth]{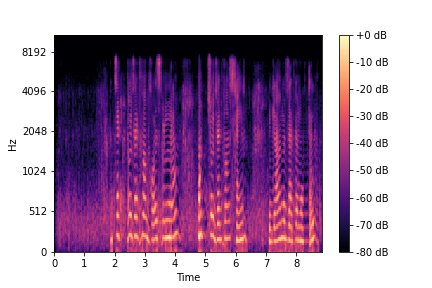}%
  \label{fig:evaluation:avgPrice1}%
}
\subfloat[Mel-spectrogram of original \\ speech considered for ALGAN-VC]{%
  \includegraphics[width=0.255\textwidth]{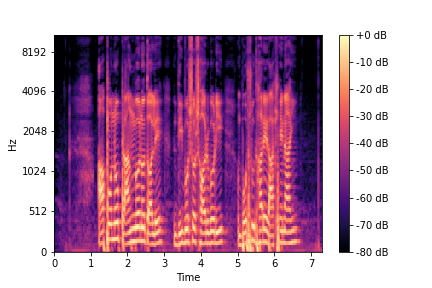}%
  \label{fig:evaluation:avgPrice2}%
}
\subfloat[Mel-spectrogram of converted \\ speech using ALGAN-VC]{%
  \includegraphics[width=0.255\textwidth]{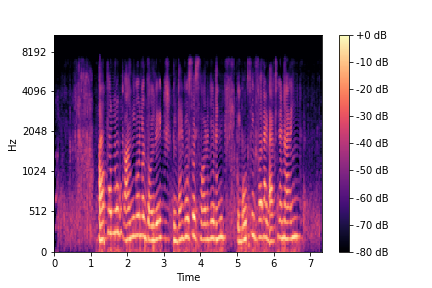}%
  \label{fig:evaluation:avgPrice3}%
}
\caption{Visualization of the Mel-spectrograms for both original and converted speech samples considered for SP-CycleGAN and ALGAN-VC.}
\label{Mel}
\end{figure*}

Finally, visual inspection is discussed with the help of Mel-spectrograms. The Mel-spectrograms of the original speech considered for SP-CycleGAN and ALGAN-VC as well as the Mel-spectrograms of the corresponding converted speech, generated by SP-CycleGAN and ALGAN-VC are shown in Fig. \ref{Mel}. 
Here original speech is considered for male speaker and the converted speech is considered for female speaker. From the visual inspection of Mel-spectrograms, the distortion of frequency components can be seen in case of the converted speech samples of SP-CycleGAN (Fig. \ref{fig:evaluation:avgPrice1}). On the other hand, the  Mel-spectrogram of the converted speech for ALGAN-VC is quite similar with the original speech. This implies better similarity of converted speech with the original speech for the ALGAN-VC model.

\section{Conclusion} \label{conclusion}
 In this paper, an adaptive learning based ALGAN-VC model is proposed for one-to-one voice conversion. The model is incorporated with three components such as DRN architecture based generator network, linear combination of $L_1$ and  $L_2$ loss functions, and a boosted learning rate strategy.The proposed model well performed the voice conversion task in terms of speaker similarity and speech quality. All the experiments are carried out over the VCC 2016, VCC 2018, VCC 2020 and our self prepared speech datasets. The experimental results of the subjective and objective evaluations revealed that our proposed model showed better performance for both mono lingual and cross lingual voice conversion as well as for intra gender and inter gender voice conversion compared to other models. Moreover, ablation study of the proposed model demonstrated the utility of $L_1$ and $L_2$ loss in the linear combination of these losses. In future, the model will be investigated in many-to-many voice conversion task. Further, speech-to-speech based and text-to-speech based GAN architectures will be developed for improving speaker similarity and speech quality of converted speech samples.
. 
\ifCLASSOPTIONcaptionsoff
  \newpage
\fi



\bibliographystyle{IEEEtran}
\bibliography{bibliography.bib}

\begin{thebibliography}{10}
\providecommand{\url}[1]{#1}
\csname url@samestyle\endcsname
\providecommand{\newblock}{\relax}
\providecommand{\bibinfo}[2]{#2}
\providecommand{\BIBentrySTDinterwordspacing}{\spaceskip=0pt\relax}
\providecommand{\BIBentryALTinterwordstretchfactor}{4}
\providecommand{\BIBentryALTinterwordspacing}{\spaceskip=\fontdimen2\font plus
\BIBentryALTinterwordstretchfactor\fontdimen3\font minus
  \fontdimen4\font\relax}
\providecommand{\BIBforeignlanguage}[2]{{%
\expandafter\ifx\csname l@#1\endcsname\relax
\typeout{** WARNING: IEEEtran.bst: No hyphenation pattern has been}%
\typeout{** loaded for the language `#1'. Using the pattern for}%
\typeout{** the default language instead.}%
\else
\language=\csname l@#1\endcsname
\fi
#2}}
\providecommand{\BIBdecl}{\relax}
\BIBdecl

\bibitem{speech-disorder}
Y.~{Zhao}, M.~{Kuruvilla-Dugdale}, and M.~{Song}, ``Voice conversion for
  persons with amyotrophic lateral sclerosis,'' \emph{IEEE Journal of
  Biomedical and Health Informatics}, vol.~24, no.~10, pp. 2942--2949, 2020.

\bibitem{tv-dubbing}
J.~{Matoušek}, Z.~{Hanzlíček}, D.~{Tihelka}, and M.~{Méner}, ``Automatic
  dubbing of tv programmes for the hearing impaired,'' in \emph{IEEE 10th
  International conference on signal processing proceedings}, Beijing, China,
  24-28 Oct 2010, pp. 589--592.

\bibitem{singing-voice-conversion}
B.~{Sisman}, K.~{Vijayan}, M.~{Dong}, and H.~{Li}, ``Singan: Singing voice
  conversion with generative adversarial networks,'' in \emph{2019 Asia-Pacific
  Signal and Information Processing Association Annual Summit and Conference
  (APSIPA ASC)}, 2019, pp. 112--118.

\bibitem{review-paper-1}
B.~{Sisman}, J.~{Yamagishi}, S.~{King}, and H.~{Li}, ``An overview of voice
  conversion and its challenges: From statistical modeling to deep learning,''
  \emph{IEEE/ACM Transactions on Audio, Speech, and Language Processing},
  vol.~29, pp. 132--157, 2021.

\bibitem{speech-synthesis-review}
A.~{Breen}, ``Speech synthesis models: a review,'' \emph{Electronics
  Communication Engineering Journal}, vol.~4, no.~1, pp. 19--31, 1992.

\bibitem{disadvantages-vc}
K.~Kuligowska, P.~Kisielewicz, and A.~Włodarz, ``Speech synthesis systems:
  disadvantages and limitations,'' \emph{International Journal of Engineering
  and Technology}, vol.~7, pp. 234--239, 2018.

\bibitem{gan-ian-goodfellow}
I.~Goodfellow, J.~Pouget-Abadie, M.~Mirza, B.~Xu, D.~Warde-Farley, S.~Ozair,
  A.~Courville, and Y.~Bengio, ``Generative adversarial nets,'' in
  \emph{Advances in Neural Information Processing Systems 27}, 2014, pp.
  2672--2680.

\bibitem{deep-learning-1}
S.~{Lee}, B.~{Ko}, K.~{Lee}, I.~C. {Yoo}, and D.~{Yook}, ``Many-to-many voice
  conversion using conditional cycle-consistent adversarial networks,'' in
  \emph{ICASSP 2020 - 2020 IEEE International Conference on Acoustics, Speech
  and Signal Processing (ICASSP)}, 2020, pp. 6279--6283.

\bibitem{deep-learning-2}
Z.~Chen, A.~Rosenberg, Y.~Zhang, G.~Wang, B.~Ramabhadran, and P.~J. Moreno,
  ``{Improving Speech Recognition Using GAN-Based Speech Synthesis and
  Contrastive Unspoken Text Selection},'' in \emph{Proc. Interspeech 2020},
  2020, pp. 556--560.

\bibitem{deep-learning-3}
H.~{Kameoka}, T.~{Kaneko}, K.~{Tanaka}, and N.~{Hojo}, ``Stargan-vc:
  non-parallel many-to-many voice conversion using star generative adversarial
  networks,'' in \emph{2018 IEEE Spoken Language Technology Workshop (SLT)},
  2018, pp. 266--273.

\bibitem{deep-learning-5}
M.~{Patel}, M.~{Parmar}, S.~{Doshi}, N.~J. {Shah}, and H.~A. {Patil}, ``Novel
  adaptive generative adversarial network for voice conversion,'' in \emph{2019
  Asia-Pacific Signal and Information Processing Association Annual Summit and
  Conference (APSIPA ASC)}, 2019, pp. 1273--1281.

\bibitem{deep-learning-4}
Z.~Luo, J.~Chen, T.~Takiguchi, and Y.~Ariki, ``Neutral-to-emotional voice
  conversion with cross-wavelet transform f0 using generative adversarial
  networks,'' \emph{APSIPA Transactions on Signal and Information Processing},
  vol.~8, 2019.

\bibitem{deep-learning-6}
C.~chieh Yeh, P.~chun Hsu, J.-C. Chou, H.~yi~Lee, and L.~Lee, ``Rhythm-flexible
  voice conversion without parallel data using cycle-gan over phoneme
  posteriorgram sequences,'' \emph{2018 IEEE Spoken Language Technology
  Workshop (SLT)}, pp. 274--281, 2018.

\bibitem{deep-learning-7}
S.~Park, D.~Kim, and M.~chul Joe, ``Cotatron: Transcription-guided speech
  encoder for any-to-many voice conversion without parallel data,'' in
  \emph{INTERSPEECH}, 2020.

\bibitem{deep-learning-8}
M.~Pasini, ``Melgan-vc: Voice conversion and audio style transfer on
  arbitrarily long samples using spectrograms,'' \emph{ArXiv}, vol.
  abs/1910.03713, 2019.

\bibitem{deep-learning-9}
R.~{Yamamoto}, E.~{Song}, and J.~M. {Kim}, ``Parallel wavegan: A fast waveform
  generation model based on generative adversarial networks with
  multi-resolution spectrogram,'' in \emph{ICASSP 2020 - 2020 IEEE
  International Conference on Acoustics, Speech and Signal Processing
  (ICASSP)}, 2020, pp. 6199--6203.

\bibitem{deep-learning-10}
W.~{Zhao}, W.~{Wang}, Y.~{Sun}, and T.~{Tang}, ``Singing voice conversion based
  on wd-gan algorithm,'' in \emph{2019 IEEE 4th Advanced Information
  Technology, Electronic and Automation Control Conference (IAEAC)}, vol.~1,
  2019, pp. 950--954.

\bibitem{deep-learning-12}
H.~{Kameoka}, T.~{Kaneko}, K.~{Tanaka}, and N.~{Hojo}, ``Nonparallel voice
  conversion with augmented classifier star generative adversarial networks,''
  \emph{IEEE/ACM Transactions on Audio, Speech, and Language Processing},
  vol.~28, pp. 2982--2995, 2020.

\bibitem{deep-learning-13}
D.~G. {Rajpura}, J.~{Shah}, M.~{Patel}, H.~{Malaviya}, K.~{Phatnani}, and H.~A.
  {Patil}, ``Effectiveness of transfer learning on singing voice conversion in
  the presence of background music,'' in \emph{2020 International Conference on
  Signal Processing and Communications (SPCOM)}, 2020, pp. 1--5.

\bibitem{deep-learning-14}
R.~Daher, M.~K. Zein, J.~E. Zini, M.~Awad, and D.~Asmar, ``Change your singer:
  A transfer learning generative adversarial framework for song to song
  conversion,'' \emph{2020 International Joint Conference on Neural Networks
  (IJCNN)}, pp. 1--7, 2020.

\bibitem{deep-learning-15}
T.~Li, Y.~Liu, C.~Hu, and H.~Zhao, ``Cvc: Contrastive learning for non-parallel
  voice conversion,'' \emph{ArXiv}, vol. abs/2011.00782, 2020.

\bibitem{cyclegan-vc-1}
T.~{Kaneko} and H.~{Kameoka}, ``Cyclegan-vc: Non-parallel voice conversion
  using cycle-consistent adversarial networks,'' in \emph{2018 26th European
  Signal Processing Conference (EUSIPCO)}, 2018, pp. 2100--2104.

\bibitem{cyclegan-vc-2}
T.~{Kaneko}, H.~{Kameoka}, K.~{Tanaka}, and N.~{Hojo}, ``Cyclegan-vc2: Improved
  cyclegan-based non-parallel voice conversion,'' in \emph{ICASSP 2019 - 2019
  IEEE International Conference on Acoustics, Speech and Signal Processing
  (ICASSP)}, 2019, pp. 6820--6824.

\bibitem{cyclegan-vc-3}
T.~Kaneko, H.~Kameoka, K.~Tanaka, and N.~Hojo, ``Cyclegan-vc3: Examining and
  improving cyclegan-vcs for mel-spectrogram conversion,'' in \emph{Proceedings
  of the Annual Conference of the International Speech Communication
  Association}, 2020.

\bibitem{cyclegan-vc-4}
S.~{Seshadri}, L.~{Juvela}, J.~{Yamagishi}, O.~{Räsänen}, and P.~{Alku},
  ``Cycle-consistent adversarial networks for non-parallel vocal effort based
  speaking style conversion,'' in \emph{ICASSP 2019 - 2019 IEEE International
  Conference on Acoustics, Speech and Signal Processing (ICASSP)}, 2019, pp.
  6835--6839.

\bibitem{cyclegan-vc-5}
X.~Jia, J.~Tai, H.~Zhou, Y.~Li, W.~Zhang, H.~Du, and Q.~Huang, ``Et-gan:
  Cross-language emotion transfer based on cycle-consistent generative
  adversarial networks,'' in \emph{ECAI}, 2020.

\bibitem{cyclegan-vc-7}
Z.~{Du}, K.~{Zhou}, B.~{Sisman}, and H.~{Li}, ``Spectrum and prosody conversion
  for cross-lingual voice conversion with cyclegan,'' in \emph{2020
  Asia-Pacific Signal and Information Processing Association Annual Summit and
  Conference (APSIPA ASC)}, 2020, pp. 507--513.

\bibitem{cyclegan-vc-8}
R.~{Ferro}, N.~{Obin}, and A.~{Roebel}, ``Cyclegan voice conversion of spectral
  envelopes using adversarial weights,'' in \emph{2020 28th European Signal
  Processing Conference (EUSIPCO)}, 2021, pp. 406--410.

\bibitem{cyclegan-vc-9}
E.~Hosseini-Asl, Y.~Zhou, C.~Xiong, and R.~Socher, ``A multi-discriminator
  cyclegan for unsupervised non-parallel speech domain adaptation,'' in
  \emph{INTERSPEECH}, 2018.

\bibitem{wave-cyclegan-2}
K.~Tanaka, H.~Kameoka, T.~Kaneko, and N.~Hojo, ``Wavecyclegan2: Time-domain
  neural post-filter for speech waveform generation,'' \emph{ArXiv}, vol.
  abs/1904.02892, 2019.

\bibitem{childrens-gan}
N.~Jia, C.~Zheng, and W.~Sun, ``Speech synthesis of children's reading based on
  {CycleGAN} model,'' \emph{Journal of Physics: Conference Series}, vol. 1607,
  p. 012046, aug 2020.

\bibitem{cyclegan-vc-100}
K.~Yasuda, R.~Orihara, Y.~Sei, Y.~Tahara, and A.~Ohsuga, ``Transforming the
  emotion in speech using a generative adversarial network,'' in \emph{ICAART},
  2019.

\bibitem{cyclegan-vc-101}
A.~{Mathur}, A.~{Isopoussu}, F.~{Kawsar}, N.~{Berthouze}, and N.~D. {Lane},
  ``Mic2mic: Using cycle-consistent generative adversarial networks to overcome
  microphone variability in speech systems,'' in \emph{2019 18th ACM/IEEE
  International Conference on Information Processing in Sensor Networks
  (IPSN)}, 2019, pp. 169--180.

\bibitem{cyclegan-vc-102}
R.~Shankar, J.~Sager, and A.~Venkataraman, ``{Non-Parallel Emotion Conversion
  Using a Deep-Generative Hybrid Network and an Adversarial Pair
  Discriminator},'' in \emph{Proc. Interspeech 2020}, 2020, pp. 3396--3400.

\bibitem{cyclegan-vc-103}
C.~{Wang} and Y.~B. {YU}, ``Cyclegan-vc-gp: Improved cyclegan-based
  non-parallel voice conversion,'' in \emph{2020 IEEE 20th International
  Conference on Communication Technology (ICCT)}, 2020, pp. 1281--1284.

\bibitem{vcc-2016-audio-data}
T.~Toda, L.-H. Chen, D.~Saito, F.~Villavicencio, M.~Wester, Z.~Wu, and
  J.~Yamagishi, ``The voice conversion challenge 2016,'' in \emph{Proc.
  INTERSPEECH}, 2016, pp. 1632--1636.

\bibitem{vcc-2018-audio-data}
T.~Kinnunen, J.~Lorenzo-Trueba, J.~Yamagishi, T.~Toda, D.~Saito,
  F.~Villavicencio, and Z.~Ling, ``A spoofing benchmark for the 2018 voice
  conversion challenge: leveraging from spoofing countermeasures for speech
  artifact assessment,'' in \emph{Proc. Odyssey 2018}, 2018, pp. 187--194.

\bibitem{vcc-2020-audio-data}
Z.~Yi, W.-C. Huang, X.~Tian3, J.~Yamagishi, R.~K. Das, T.~Kinnunen, Z.~Ling,
  and T.~Toda, ``Voice conversion challenge 2020 intra lingual semi-parallel
  and cross-lingual voice conversion,'' in \emph{Joint Workshop for the
  Blizzard Challenge and Voice Conversion Challenge 2020}, 2020.

\bibitem{Review-paper-on-GAN}
H.~Alqahtani, M.~Kavakli-Thorne, and D.~G. Kumar~Ahuja, ``Applications of
  generative adversarial networks (gans): An updated review,'' \emph{Archives
  of Computational Methods in Engineering}, vol.~28, 12 2019.

\bibitem{lOSS1}
X.~{Mao}, Q.~{Li}, H.~{Xie}, R.~Y.~K. {Lau}, Z.~{Wang}, and S.~P. {Smolley},
  ``Least squares generative adversarial networks,'' in \emph{2017 IEEE
  International Conference on Computer Vision (ICCV)}, 2017, pp. 2813--2821.

\bibitem{lOSSfunc2}
M.~Arjovsky, S.~Chintala, and L.~Bottou, ``Wasserstein generative adversarial
  networks,'' in \emph{Proceedings of the 34th International Conference on
  Machine Learning - Volume 70}, ser. ICML'17.\hskip 1em plus 0.5em minus
  0.4em\relax JMLR.org, 2017, p. 214–223.

\bibitem{loss-3}
S.~Nowozin, B.~Cseke, and R.~Tomioka, ``F-gan: Training generative neural
  samplers using variational divergence minimization,'' in \emph{Proceedings of
  the 30th International Conference on Neural Information Processing Systems},
  ser. NIPS'16.\hskip 1em plus 0.5em minus 0.4em\relax Red Hook, NY, USA:
  Curran Associates Inc., 2016, p. 271–279.

\bibitem{CycleGAN2017}
J.-Y. Zhu, T.~Park, P.~Isola, and A.~A. Efros, ``Unpaired image-to-image
  translation using cycle-consistent adversarial networks,'' in \emph{Computer
  Vision (ICCV), 2017 IEEE International Conference on}, 2017.

\bibitem{world}
M.~Morise, F.~YOKOMORI, and K.~Ozawa, ``World: A vocoder-based high-quality
  speech synthesis system for real-time applications,'' \emph{IEICE
  Transactions on Information and Systems}, vol. E99.D, pp. 1877--1884, 07
  2016.

\bibitem{PML}
G.~{Degottex}, P.~{Lanchantin}, and M.~{Gales}, ``A log domain pulse model for
  parametric speech synthesis,'' \emph{IEEE/ACM Transactions on Audio, Speech,
  and Language Processing}, vol.~26, no.~1, pp. 57--70, 2018.

\bibitem{Griffin-Lim}
D.~{Griffin} and {Jae Lim}, ``Signal estimation from modified short-time
  fourier transform,'' in \emph{ICASSP '83. IEEE International Conference on
  Acoustics, Speech, and Signal Processing}, vol.~8, 1983, pp. 804--807.

\bibitem{Resnet}
K.~He, X.~Zhang, S.~Ren, and J.~Sun, ``Deep residual learning for image
  recognition,'' in \emph{2016 IEEE Conference on Computer Vision and Pattern
  Recognition (CVPR)}, 2016.

\bibitem{Resnets-for-speech-enhancements}
T.~A. {Hsieh}, H.~M. {Wang}, X.~{Lu}, and Y.~{Tsao}, ``Wavecrn: An efficient
  convolutional recurrent neural network for end-to-end speech enhancement,''
  \emph{IEEE Signal Processing Letters}, vol.~27, pp. 2149--2153, 2020.

\bibitem{Resnets-for-speech-enhancements2}
D.~Gonz{\'a}lez, J.~Llombart, A.~Miguel, and L.~Vicente, ``Deep speech
  enhancement for reverberated and noisy signals using wide residual
  networks,'' \emph{ArXiv}, vol. abs/1901.00660, 2019.

\bibitem{DenseNet-new}
G.~{Huang}, Z.~{Liu}, L.~{Van Der Maaten}, and K.~Q. {Weinberger}, ``Densely
  connected convolutional networks,'' in \emph{2017 IEEE Conference on Computer
  Vision and Pattern Recognition (CVPR)}, 2017, pp. 2261--2269.

\bibitem{GAN-LOSS-FUNCTION}
Z.~{Pan}, W.~{Yu}, B.~{Wang}, H.~{Xie}, V.~S. {Sheng}, J.~{Lei}, and
  S.~{Kwong}, ``Loss functions of generative adversarial networks (gans):
  Opportunities and challenges,'' \emph{IEEE Transactions on Emerging Topics in
  Computational Intelligence}, vol.~4, no.~4, pp. 500--522, 2020.

\bibitem{Adaptive-activation-function-1}
L.~Nanni, A.~Lumini, S.~Ghidoni, and G.~Maguolo, ``Stochastic selection of
  activation layers for convolutional neural networks,'' \emph{Sensors},
  vol.~20, no.~6, 2020.

\bibitem{Sppech-L1}
A.~{Pandey} and D.~{Wang}, ``On adversarial training and loss functions for
  speech enhancement,'' in \emph{2018 IEEE International Conference on
  Acoustics, Speech and Signal Processing (ICASSP)}, 2018, pp. 5414--5418.

\bibitem{Speech-L1-1}
S.~Pascual, A.~Bonafonte, and J.~Serr{\`a}, ``Segan: Speech enhancement
  generative adversarial network,'' \emph{INTERSPEECH 2017}, 2017.

\bibitem{L2}
K.-S. Lee, ``Voice conversion using a perceptual criterion,'' \emph{Applied
  Sciences}, vol.~10, no.~8, 2020.

\bibitem{Activation-Function-Ensambling}
X.~Wang, J.~Jiang, M.~Gao, Z.~Liu, and C.~Zhao, ``{Activation ensemble
  generative adversarial network transfer learning for image classification},''
  \emph{Journal of Electronic Imaging}, vol.~30, no.~1, pp. 1 -- 15, 2021.

\bibitem{learning-rate-boosting}
J.~Park, D.~Yi, and S.~Ji, ``A novel learning rate schedule in optimization for
  neural networks and it’s convergence,'' \emph{Symmetry}, vol.~12, no.~4,
  2020.

\bibitem{2016}
T.~Toda, L.-H. Chen, D.~Saito, F.~Villavicencio, M.~Wester, Z.~Wu, and
  J.~Yamagishi, ``The voice conversion challenge 2016,'' in \emph{Interspeech
  2016}, 2016, pp. 1632--1636.

\bibitem{2018}
J.~Lorenzo-Trueba, J.~Yamagishi, T.~Toda, D.~Saito, F.~Villavicencio,
  T.~Kinnunen, and Z.~Ling, ``The voice conversion challenge 2018: Promoting
  development of parallel and nonparallel methods,'' in \emph{Proc. Odyssey
  2018 The Speaker and Language Recognition Workshop}, 2018, pp. 195--202.

\bibitem{2020}
Z.~Yi, W.-C.~H. Chen, X.~Tian, J.~Yamagishi, R.~K. Das, T.~Kinnunen, Z.~Ling5,
  and T.~Toda, ``The voice conversion challenge 2016,'' in \emph{Joint Workshop
  for the Blizzard Challenge and Voice Conversion Challenge 2020}, 2016.

\bibitem{Gaussian}
K.~{Liu}, J.~{Zhang}, and Y.~{Yan}, ``High quality voice conversion through
  phoneme-based linear mapping functions with straight for mandarin,'' in
  \emph{Fourth International Conference on Fuzzy Systems and Knowledge
  Discovery (FSKD 2007)}, vol.~4, 2007, pp. 410--414.

\bibitem{APS}
Y.~Ohtani, T.~Toda, H.~Saruwatari, and K.~Shikano, ``Maximum likelihood voice
  conversion based on gmm with straight mixed excitation,'' in
  \emph{INTERSPEECH}, 2006.

\bibitem{STyle-transfer}
J.~J., A.~A., and F.-F. L., ``Perceptual losses for real-time style transfer
  and super-resolution,'' in \emph{ECCV 2016}, 2016, pp. 694--711.

\bibitem{IN}
D.~Ulyanov, A.~Vedaldi, and V.~Lempitsky, ``Instance normalization: The missing
  ingredient for fast stylization,'' \emph{ArXiv}, vol. abs/1607.08022, 2016.

\bibitem{cvpr}
W.~Shi, J.~Caballero, F.~Husz{\'a}r, J.~Totz, A.~P. Aitken, R.~Bishop,
  D.~Rueckert, and Z.~Wang, ``Real-time single image and video super-resolution
  using an efficient sub-pixel convolutional neural network,'' \emph{2016 IEEE
  Conference on Computer Vision and Pattern Recognition (CVPR)}, pp.
  1874--1883, 2016.

\bibitem{2DCNN}
T.~{Kaneko}, H.~{Kameoka}, N.~{Hojo}, Y.~{Ijima}, K.~{Hiramatsu}, and
  K.~{Kashino}, ``Generative adversarial network-based postfilter for
  statistical parametric speech synthesis,'' in \emph{2017 IEEE International
  Conference on Acoustics, Speech and Signal Processing (ICASSP)}, 2017, pp.
  4910--4914.

\bibitem{abx-test}
A.~{Rajpal}, N.~J. {Shah}, M.~{Zaki}, and H.~A. {Patil}, ``Quality assessment
  of voice converted speech using articulatory features,'' in \emph{2017 IEEE
  International Conference on Acoustics, Speech and Signal Processing
  (ICASSP)}, 2017, pp. 5515--5519.

\bibitem{mos}
Y.~{Li}, {Zhang Ling-hua}, and {Ding Hui}, ``Nonparallel voice conversion based
  on phoneme classification and eigenvoices,'' in \emph{2010 IEEE 12th
  International Conference on Communication Technology}, 2010, pp. 662--665.

\bibitem{resemble-ai}
\BIBentryALTinterwordspacing
Resemble-Ai, ``resemble-ai/resemblyzer.'' [Online]. Available:
  \url{https://github.com/resemble-ai/Resemblyzer}
\BIBentrySTDinterwordspacing

\end{thebibliography}
%

%








\end{document}